\documentclass[journal]{IEEEtranTCOM}

\usepackage{graphicx}

\newtheorem{theorem}{\bf Theorem}

\begin{document}


\title{A Generalized Channel Coding Theory for Distributed Communication}

\author{Jie Luo,~\IEEEmembership{Senior Member, IEEE}
\thanks{The author is with the Electrical and Computer Engineering Department, Colorado State University, Fort Collins, CO 80523. E-mail: rockey@engr.colostate.edu. }
\thanks{This work was supported by the National Science Foundation under Grants CCF-1016985, CNS-1116134 and CCF-1420608. Any opinions, findings, and conclusions or recommendations expressed in this paper are those of the authors and do not necessarily reflect the views of the National Science Foundation.}
\thanks{Part of the results of this paper have been presented at the 2014 IEEE International Conference on Communications \cite{ref Luo14}.}
}




\maketitle

\begin{abstract}
This paper presents generalized channel coding theorems for a time-slotted distributed communication system where a transmitter-receiver pair is communicating in parallel with other transmitters. Assume that the channel code of each transmitter is chosen arbitrarily in each time slot. The coding choice of a transmitter is denoted by a code index parameter, which is known neither to other transmitters nor to the receiver. Fundamental performance limitation of the system is characterized using an achievable region defined in the space of the code index vectors. As the codeword length is taken to infinity, for all code index vectors inside the region, the receiver will decode the message reliably, while for all code index vectors outside the region, the receiver will report a collision reliably. A generalized system error performance measure is defined as the weighted sum of probabilities of different types of communication error events. Assume that the receiver chooses an ``operation region" and intends to decode the message if the code index vector is inside the operation region. Achievable bounds on the tradeoff between the operation region and the generalize error performance measure are obtained under the assumption of a finite codeword length.
\end{abstract}


\section{Introduction}
Classical multiuser channel coding theory assumes that transmitters and receivers involved in a communication party should jointly determine their channel codes before message transmission \cite{ref Cover05}. Channel code design includes the full determination of communication parameters such as information rate and transmission power. Under the assumption of long message transmission over a stationary channel, overhead of joint channel coding is ignored due to its negligible weight in the asymptotic communication efficiency. Although channel coding theory has been extended to non-asymptotic cases with finite codeword lengths \cite{ref Gallager65}\cite{ref Polyanskiy10}, as well as to communication systems that involve random transmission activities \cite{ref Hui84}\cite{ref Massey85}\cite{ref Minero12}, the majority of these extensions still inherited the basic assumption that continuous transmission of encoded long messages should dominate the communication process. Consequently, joint channel coding among transmitters and receivers in the communication party, which is referred to as ``coordinated communication" in this paper, remains a fundamental assumption in most of the classical channel coding problem formulations.

The rapid expansion and the growing complexity of wireless networks have generated communication scenarios that are quite different from those envisioned in the classical channel coding theory. Due to the dynamic nature of networking activities, users in a communication network often have bursty short messages that must be disseminated in a timely manner. Full communication coordination among a large group of wireless users over a long time duration is often expensive or infeasible in the sense of excessive overhead. The high cost or infeasibility of coordinated communication can be caused by a wide range of factors including computational complexity, communication complexity, privacy issues, robustness requirements, or simply the lack of a universal coordination protocol. Consequently, a significant proportion of data transmissions in current wireless networks are carried out using distributed communication protocols where transmitters determine their channel codes and communication parameters individually. Although coding redundancy is still needed to improve communication reliability, distributed communication has been largely ignored in the classical channel coding literature and its fundamental performance limits are far from being well understood.

Because distributed communication arises from network applications, modularized network architecture is an important aspect that needs to be carefully considered in its channel coding problem formulations. Take the layered network architecture \cite{ref Bertskas92} for example. Due to the basic assumption that communication optimization should be carried out at the physical layer, data link layer only determines when and whether a (link-layer) user should transmit a packet to the corresponding receiver. In other words, transmission options of a data link layer user are binary (transmitting/idling). On the other hand, the layered architecture also requires that physical layer should verify whether a planned message transmission can be made reliable. When reliable message delivery cannot be achieved, physical layer receiver should report a collision (or outage) to the link layer rather than confusing the upper layer with unreliable message estimates. Such a function is termed ``collision detection" in this paper. When physical layer users are fully coordinated, collision detection is done jointly by the involved transmitters and receivers at the stage of communication planning.

In a distributed network, communication coordination and optimization cannot be done fully at the physical layer. Consequently, data link layer must share the responsibility of communication adaptation to improve medium access control performance. Unfortunately, with the binary transmitting/idling options assumed in the current network architecture, choices available for communication adaptation at the data link layer are very limited. Although medium access control is beyond the scope of our channel coding research, enabling an increased number of transmission options at the data link layer to support advanced communication adaptation is a key motivation behind the channel coding problem formulation to be presented in this paper. Furthermore, due to lack of global communication information, physical layer transmitters in a distributed network may not have the capability to know whether a transmitted message can be received reliably or not. Consequently, collision detection becomes an explicit responsibility of each physical layer receiver.

In \cite{ref Luo12}, we proposed a new channel coding model for time-slotted random multiple access communication systems. We focused on communication and coding within one time-slot or one packet. Each transmitter is equipped with a randomly generated codebook that supports multiple communication rate options \cite{ref Luo12}. Communication rate of each transmitter is determined arbitrarily, with the rate information being shared neither among the transmitters nor with the receiver. An achievable rate region was defined in a sense explained in \cite{ref Luo12}. We showed that the achievable rate region coincides with the Shannon information rate region of the multiple access channel without a convex hull operation \cite{ref Luo12}. The asymptotic result was then extended in \cite{ref Wang12} to a rate and error probability tradeoff bound under the assumption of a finite codeword length, and in \cite{ref Wang11} to random access communication over a compound channel. Compared with a classical channel coding model, the system models of \cite{ref Luo12}\cite{ref Wang12}\cite{ref Wang11} extended the definition of ``communication error" from its classical meaning of erroneous message decoding to the new meaning of failing to give the expected outcome, whose definition should be specified by the physical layer module. By adding reliable collision report into the list of expected communication outcomes under certain conditions, the extension on communication error definition enabled the relaxation of joint channel coding requirement at the transmitters, and this consequently established a bridge toward developing rigorous channel coding theorems for distributed communication systems. Recently, it was shown in \cite{ref Farkas14} that, with the help of constant composition codes at the transmitters and maximum mutual information decoder \cite{ref Csiszar11} at the receiver, decoding error exponent of the random multiple access system can be further improved to possess an interesting decoupling property \cite{ref Farkas14}. Early investigation on the impact of the new channel coding framework on medium access control of distributed wireless networks was reported in \cite{ref Tang14}.

This paper further extends the results of \cite{ref Luo12}\cite{ref Wang12}\cite{ref Wang11} in a wide range of aspects. The extensions are briefly outlined below and are explained in detail in the paper. First, we consider a general distributed communication scenario where a pair of transmitter and receiver are communicating in parallel with other transmitters. The receiver only cares about decoding the message of its own transmitter although the messages of other transmitters can be decoded if necessary. Second, we assume that each transmitter can choose its own channel code without sharing such information either with other transmitters or with the receiver. Different channel coding choices imply different values of communication parameters that include but are not limited to the communication rate. Third, we introduce ``interfering users" whose codebook information is only partially known at the receiver. We show that, in addition to modeling remote interfering transmitters, interfering users can also be used to model the impact of a compound channel where the channel state is not precisely known at the receiver. Fourth, we introduce a generalized error performance measure that allows the system to target different exponential probability scaling laws for different types of communication error events. Consequently, a wide range of error performance objectives can be considered when deriving the performance tradeoff bound in the case of a finite codeword length. Fifth, we show that depending on the definition of communication error events, collision detection part of the channel coding problem formulation can be ill-posed and can lead to a performance bottleneck in the technical results, but may not necessarily reflect the actual design objective of the distributed communication system. Discussions on variations of communication error definition and their impact on the error performance of the system are presented. Finally, approaches that reduce the computational complexity of the channel coding schemes are also discussed.

\section{Problem Formulation}
\label{SectionII}
Consider a distributed communication system where one transmitter and receiver pair is communicating in parallel with $K+M-1$ other transmitters, where $K>0$ and $M\ge 0$. The transmitters are indexed from $1$ to $K+M$ with transmitter $1$ being paired with the receiver. Time is slotted with the length of each slot equaling $N$ symbol durations, which is also the length of a packet or a codeword. We assume that channel coding is applied only within each time slot. The channel is characterized by a conditional distribution $P_{Y|X_{1},\cdots,X_{K+M}}$ where, for $k \in \{1,\cdots, K+M\}$, $X_k \in \mathcal X$ is the channel input symbol of user $k$ with $\mathcal X$ being the finite input alphabet, and $Y \in \mathcal Y$ is the channel output symbol with $\mathcal Y$ being the finite output alphabet\footnote{Coding results presented in this paper can be extended to systems with continuous input/output alphabets in the same way as in the classical channel coding theory \cite{ref Cover05}. The basic idea is to quantize the input/output symbols and to consider quantization error bounds when deriving the corresponding coding theorems.}. Assume that at the beginning of a time slot, each transmitter, say transmitter $k$, chooses an arbitrary\footnote{Here ``arbitrary" means that the decision is made randomly with its statistical information possibly unavailable at the physical layer.} channel code, denoted by a code index parameter $g_k \in \mathcal G_k$ where $\mathcal G_k=\{g_{k1},\cdots,g_{k|\mathcal G_k|}\}$ is a finite alphabet with cardinality $|\mathcal G_k|$. The code index parameter is shared neither among the transmitters nor with the receiver. Let $r_k(g_k)$ be the communication rate, in nats per symbol, corresponding to channel code $g_k$ of transmitter $k$. The transmitter encodes $Nr_k$ number of data nats, denoted by a message $w_k$, into a packet (codeword) of $N$ symbols. We assume a random coding scheme where codeword symbols of each code are generated i.i.d. according to a pre-determined distribution. A mathematical specification of the random coding scheme, which is revised from a similar presentation originally introduced in \cite{ref Shamai07}, is given in the following. For all $k\in\{1,\cdots,K+M\}$, transmitter $k$ is equipped with a codebook library $\mathcal{L}_k = \{\mathcal{C}_{k\theta_k}: \theta_k \in \Theta_k\}$ in which codebooks are indexed by a set $\Theta_k$. Each codebook has $|\mathcal G_k|$ classes of codewords. Each class of codewords is termed a code. The $i^{th}$ ($i \in \{1,\cdots, |\mathcal G_k|\}$) code has $\lfloor e^{Nr_{ki}} \rfloor$ codewords each containing $N$ symbols. Here $r_{ki}$ is the communication rate of code $g_{ki}$. Note that in this coding scheme, each codeword in the codebook is mapped to a message and code index pair $(w_k,g_k)$. Let $\mathcal{C}_{k\theta_k}(w_k,g_k)_j$ be the $j^{th}$ symbol of the codeword corresponding to message and code index pair $(w_k,g_k)$ in codebook $\mathcal{C}_{k\theta_k}$. Transmitter $k$ first selects the codebook by generating $\theta_k$ according to a distribution $\vartheta_k$ such that the random variables $X_{(w_k,g_k),j}: \theta_k \rightarrow \mathcal{C}_{k\theta_k}(w_k,g_k)_j$ are i.i.d. according to an input distribution $P_{X|g_k}$\footnote{Note that input distributions of different codes can be different.}. Codebook $\mathcal{C}_{k\theta_k}$ is then used to map $(w_k,g_k)$ into a codeword, denoted by $\mbox{\boldmath$x$}_{(w_k,g_k)}$. After encoding, codewords of all transmitters are sent to the receiver over the discrete memoryless channel.

We use a bold font vector variable to denote the corresponding variables of all transmitters. For example, $\mbox{\boldmath $w$}$ and $\mbox{\boldmath $g$}$ denote the messages and code indices of all transmitters. $\mbox{\boldmath $P$}_{\mbox{\scriptsize\boldmath $X$}|\mbox{\scriptsize\boldmath $g$}}$ denote the input distributions of all transmitters, etc. Given a vector variable $\mbox{\boldmath$g$}$, we use $g_k$ to denote its element corresponding to transmitter $k$. Let $\mathcal S \subset \{1,\cdots,K+M\}$ be a transmitter subset, and $\bar{\mathcal{S}}$ be its complement. We use $\mbox{\boldmath $g$}_{\mathcal S}$ to denote the vector that is extracted from $\mbox{\boldmath $g$}$ with only elements corresponding to transmitters in $\mathcal S$.

We categorize transmitters with indices $\{1,\cdots, K\}$ as ``regular users" and other transmitters as ``interfering users". For each regular user (transmitter) $k\in \{1,\cdots, K\}$, we assume that the receiver knows the randomly selected codebook $\mathcal{C}_{k\theta_k}$. Codebook information can be conveyed by sharing the random codebook generation algorithm with the receiver. Note that this does not imply significant online information exchange. For example, assume that codebook library $\mathcal{L}_k$ of user $k$ is specified in the physical layer protocol while codebook index $\theta_k$ is generated using user identity and time information. So long as the receiver expects transmitter $k$ in the area and is also synchronized in time with the transmitter, no further online information exchange is needed for the receiver to generate $\mathcal{C}_{k\theta_k}$. For each interfering user $k\in \{K+1, \cdots, K+M\}$, we assume that the receiver only knows the set of input distributions $\{P_{X|g_k}|g_k \in G_k\}$, but not the codebook $\mathcal{C}_{k\theta_k}$. In other words, messages of the interfering users are not decodable at the receiver. There are two reasons why we include interfering users in the system model. First, although it is not difficult for a receiver to generate the codebook of a transmitter, for reasons such as decoding complexity constraint, the receiver may not have the capability to fully process the codebook information of all transmitters. Regarding some of the transmitters as interfering users still allows the receiver to take advantage of their input distribution information to improve coding performance. Second, interfering user can be used to model channel uncertainty at the receiver \cite{ref Wang11}. For example, to model distributed communication over a compound channel with $|\mathcal G|$ possible realizations, one can introduce an interfering user whose code index takes $|\mathcal G|$ possible values each corresponding to a channel realization. When the interfering user chooses a specific ``code", the conditional channel distribution is set to match the corresponding channel realization. Coding theorems derived in this paper can then be applied to distributed communication with channel uncertainty at the receiver. In the latter case, we also call the interfering user a ``virtual user" since it does not represent an actual transmitter in the system\footnote{An example of channel coding analysis for random access communication over a compound channel can be found in \cite{ref Wang11}.}.

We assume that the receiver is only interested in decoding the message of transmitter $1$, although the messages of other regular users can be decoded if necessary. As explained in \cite{ref Luo12}\cite{ref Wang11}, whether the message of transmitter $1$ can be decoded reliably or not depends on the coding choices of all transmitters. We assume that, before packet transmission, the receiver pre-determines an ``operation region'' $\mathcal R$, defined in the space of the code index vectors. Determination of the operation region $\mathcal R$ depends on the performance objective of the receiver, which will be discussed later. Let $\mbox{\boldmath$g$}$ be the actual code index vector with the corresponding rate vector being $\mbox{\boldmath$r$}$. We assume that the receiver {\it intends} to decode the message of transmitter $1$ if $\mbox{\boldmath$g$} \in \mathcal R$. The receiver {\it intends} to report a collision for transmitter $1$ if $\mbox{\boldmath$g$} \notin \mathcal R$. Note that $\mbox{\boldmath$g$}$ is unknown to the receiver. In each time slot, upon receiving the channel output symbols $\mbox{\boldmath$y$}$, the receiver estimates the code index vector, denoted by $\hat{\mbox{\boldmath$g$}}$, for all transmitters. The receiver outputs the estimated message and code index of transmitter $1$, denoted by $(\hat{w}_1, \hat{g}_1)$, if  $\hat{\mbox{\boldmath$g$}} \in \mathcal R$ and a pre-determined decoding error probability requirement can be met. Otherwise, the receiver reports a collision for transmitter $1$.

Given the operation region $\mathcal R$, and conditioned on $\mbox{\boldmath$g$}$ and $\mbox{\boldmath$w$}$ being the actual code index and message vectors, communication error probability as a function of $\mbox{\boldmath$g$}$ is defined as follows.
\begin{equation}\label{MC-DecodingErrorDef}
P_e({\mbox{\boldmath$g$}}) = \left\{\begin{array}{l}\max_{\mbox{\scriptsize\boldmath$w$}}Pr \left\{ (\hat{w}_1,\hat{g}_1) \neq (w_1,g_1)|(\mbox{\boldmath$w$},\mbox{\boldmath$g$}) \right\}, \forall \mbox{\boldmath$g$} \in \mathcal R \\ \max_{\mbox{\scriptsize\boldmath$w$}}1-Pr \left\{ \mbox{``collision''} \mbox{ or } \right.  \\ \quad   \left. (\hat{w}_1,\hat{g}_1) =(w_1,g_1)|(\mbox{\boldmath$w$},\mbox{\boldmath$g$}) \right\}  \quad \forall \mbox{\boldmath$g$} \not\in \mathcal R \end{array} \right.
\end{equation}
Note that the communication error definition given in (\ref{MC-DecodingErrorDef}) is slightly different from the one given in \cite{ref Luo12}\cite{ref Wang12}. More specifically, when $\mbox{\boldmath$g$} \not\in \mathcal R$, even though the receiver {\it intends} to report a collision for transmitter $1$, we also consider correct message decoding as an expected outcome. This is opposed to the communication error definition used in \cite{ref Luo12}\cite{ref Wang12}, where only collision report is regarded as the expected outcome for $\mbox{\boldmath$g$} \not\in \mathcal R$. Communication error definition of (\ref{MC-DecodingErrorDef}) is chosen based on the assumption that the primary objective of the decoder is to guarantee the reliability of its message output. In other words, whether code indices of the other transmitters are correctly detected or not is of no interest to the receiver. We will maintain this communication error definition in Sections \ref{SectionII} and \ref{SectionIII} when deriving the basic channel coding theorems. However, we want to point out that there are situations when such a seemingly natural definition deserves a careful investigation. Discussions on extensions of the communication error definition will be presented in Section \ref{SectionIV}, while explanations on the necessity of these extensions will be delayed further to Section \ref{SectionVI}.

We define the system error probability as
\begin{equation}\label{SystemError}
P_{es}=\frac{1}{\prod_1^{M+K}|{\mathcal G_k}|}\sum_{\mbox{\scriptsize\boldmath$g$}}P_e({\mbox{\boldmath$g$}}),
\end{equation}
which is the error probability if all code index vectors are chosen with an equal probability. Furthermore, let $\alpha({\mbox{\boldmath$g$}})\ge 0$ be an arbitrary function of ${\mbox{\boldmath$g$}}$, we define ``generalized error performance" of the system as
\begin{equation}\label{GenerlizedError}
\mbox{GEP}(\alpha)=\frac{\sum_{\mbox{\scriptsize\boldmath$g$}}P_e({\mbox{\boldmath$g$}}) \exp\left(-N\alpha(\mbox{\boldmath$g$})\right)}{\sum_{\mbox{\scriptsize\boldmath$g$}}\exp\left(-N\alpha({\mbox{\boldmath$g$}})\right)}.
\end{equation}
$\mbox{GEP}(\alpha)$ is the error probability of the distributed communication system if code index vector ${\mbox{\boldmath$g$}}$ is chosen with probability $\frac{\exp\left(-N\alpha(\mbox{\scriptsize \boldmath$g$})\right)}{\sum_{\mbox{\scriptsize\boldmath$g$}}\exp\left(-N\alpha({\mbox{\scriptsize\boldmath$g$}})\right)}$. Note that definitions of the system error probability and the generalized error performance measure are different from those given in our prior works \cite{ref Wang12}\cite{ref Wang11}\cite{ref Luo14}. The revisions are partially motivated by the coding results introduced in \cite{ref Farkas14} which implicitly suggested an error probability measure similar to (\ref{GenerlizedError}). Although the true prior probability of the code index vectors may not be known at the physical layer, defining a generalized error performance measure using an imposed prior probability enables the system to target different probability scaling laws for different types of communication errors. For example, when $\alpha({\mbox{\boldmath$g$}})\equiv0$, we have $\mbox{GEP}(\alpha)=P_{es}$, which means all error events are treated equally in the generalized measure. For another example, if the system only cares about message decoding but not the collision report, the objective can be reflected by setting $\alpha({\mbox{\boldmath$g$}})=0$ for ${\mbox{\boldmath$g$}}\in {\mathcal R}$ and $\alpha({\mbox{\boldmath$g$}})= \infty$ for ${\mbox{\boldmath$g$}}\not\in {\mathcal R}$.

\section{Basic Channel Coding Theorems}
\label{SectionIII}
Given a distributed communication system as described in Section \ref{SectionII}. Let us fix the coding parameters\footnote{Such as rate functions, alphabets of code indices and input distributions.} that are not functions of the codeword length. We say that an operation region is achievable if there exists a set of decoding algorithms whose system error probability converges to zero as the codeword length is taken to infinity, i.e., $\lim_{N\to \infty}P_{es}=0$. The following achievable region result is a trivial extension of the similar result given in \cite[Theorem 3]{ref Luo12}.

\begin{theorem}\label{Theorem1}
Consider the distributed communication system described in Section \ref{SectionII} with $(K+M)$ transmitters. Let $\mbox{\boldmath  $r$}$ be the communication rate vector corresponding to code index vector $\mbox{\boldmath  $g$}$, and $r_k(g_k)$ be the element of $\mbox{\boldmath  $r$}$ corresponding to transmitter $k$. The following region defined in the space of $\mbox{\boldmath  $g$}$ is achievable.
\begin{equation}
\mathcal R=\left\{\mbox{\boldmath  $g$}\left| \begin{array}{l} \forall \mathcal S \subseteq \{1, \dots, K\}, 1\in \mathcal S, \exists \tilde{\mathcal S}\subseteq \mathcal S, 1\in \tilde{\mathcal S}, \\ \mbox{such that, } \sum_{k\in \tilde{\mathcal S}}r_k(g_k)      \\ \quad     < I_{\mbox{\scriptsize \boldmath  $g$}} ( \mbox{\boldmath  $X$}_{k \in \tilde{\mathcal S}} ; Y |\mbox{\boldmath  $X$}_{k \in \{1,\cdots, K\}\setminus\mathcal S})  \end{array} \right.\right\},
\label{MutualInformationInequality1}
\end{equation}
where the mutual information term $I_{\mbox{\scriptsize \boldmath  $g$}}( \mbox{\boldmath  $X$}_{k\in \tilde{\mathcal S}} ; Y |\mbox{\boldmath  $X$}_{k \in \{1,\cdots, K\}\setminus\mathcal S})$ is computed using input distribution $\mbox{\boldmath  $P$}_{\mbox{\scriptsize \boldmath  $X$}|\mbox{\scriptsize \boldmath  $g$}}$. $\QED$
\end{theorem}

Theorem \ref{Theorem1} can be proven by following the proof of \cite[Theorem 3]{ref Luo12} with only minor revisions. Note that, although the interfering users do not show up explicitly in the expression of $\mathcal R$, their code indices do affect the the mutual information terms in (\ref{MutualInformationInequality1}).

Because both collision report and correct message decoding are included in the set of expected outcomes for $\mbox{\boldmath$g$} \not\in \mathcal R$, the following theorem follows immediately from the achievable region definition.
\begin{theorem}\label{Theorem2}
For the distributed communication system considered in Theorem \ref{Theorem1}, any subset of an achievable region is also achievable.
\end{theorem}

Next, we will consider the case when the codeword length is finite. As shown in \cite{ref Luo12}\cite{ref Wang11}, depending on the actual code index vector, which is unknown to the receiver, the receiver may need to jointly decode the messages of multiple transmitters in order to recover the message of transmitter $1$. Therefore, we will first need to analyze the performance of a ``$({\mathcal D}, {\mathcal R}_{\mathcal D})$-decoder" that targets at decoding a particular transmitter subset specified by ${\mathcal D}\subseteq \{1, \cdots, K\}$ \cite{ref Luo12}\cite{ref Wang11}. Let ${\mathcal R}_{\mathcal D}$ be the operation region of the $({\mathcal D}, {\mathcal R}_{\mathcal D})$-decoder. When the code index vector $\mbox{\boldmath  $g$}\in {\mathcal R}_{\mathcal D}$ is inside the operation region, the $({\mathcal D}, {\mathcal R}_{\mathcal D})$-decoder intends to decode the messages of all transmitters {\it and only the transmitters} in ${\mathcal D}$. When the code index vector $\mbox{\boldmath  $g$}\not\in {\mathcal R}_{\mathcal D}$ is outside the operation region, the $({\mathcal D}, {\mathcal R}_{\mathcal D})$-decoder intends to report collision for all transmitters in ${\mathcal D}$. Let $\mbox{\boldmath  $g$}$ be the actual code index vector. Let $(\hat{\mbox{\boldmath$w$}}_{\mathcal D}, \hat{\mbox{\boldmath$g$}}_{\mathcal D})$ be the decoding output. Error probability of the $({\mathcal D}, {\mathcal R}_{\mathcal D})$-decoder is defined as
\begin{eqnarray}\label{MC-DecodingErrorDefDRD}
&& P_{e{\mathcal D}}({\mbox{\boldmath$g$}})=      \nonumber \\
&& \left\{ \begin{array}{l}\max_{\mbox{\scriptsize\boldmath$w$}}Pr \left\{ (\hat{\mbox{\boldmath$w$}}_{\mathcal D}, \hat{\mbox{\boldmath$g$}}_{\mathcal D}) \neq (\mbox{\boldmath$w$}_{\mathcal D}, \mbox{\boldmath$g$}_{\mathcal D})|(\mbox{\boldmath$w$},\mbox{\boldmath$g$}) \right\},     \\ \qquad  \qquad \qquad \qquad \qquad \qquad \qquad        \forall \mbox{\boldmath$g$} \in \mathcal R_{\mathcal D} \\ \max_{\mbox{\scriptsize\boldmath$w$}}1-Pr \left\{ \mbox{``collision''} \mbox{ or } \right.     \\  \quad           \left. (\hat{\mbox{\boldmath$w$}}_{\mathcal D}, \hat{\mbox{\boldmath$g$}}_{\mathcal D}) = (\mbox{\boldmath$w$}_{\mathcal D}, \mbox{\boldmath$g$}_{\mathcal D}) |(\mbox{\boldmath$w$},\mbox{\boldmath$g$}) \right\} \quad \forall \mbox{\boldmath$g$} \not\in \mathcal R_{\mathcal D} \end{array} \right. .
\end{eqnarray}
Similar to the definition of (\ref{MC-DecodingErrorDef}), when $\mbox{\boldmath$g$} \not\in \mathcal R_{\mathcal D}$, we still regard correct message decoding as an expected outcome, as opposed to a communication error event.

Let $\alpha({\mbox{\boldmath$g$}})\ge 0$ be an arbitrary function of ${\mbox{\boldmath$g$}}$, the generalized error performance measure is defined by
\begin{equation}\label{GEPDRD}
\mbox{GEP}_{\mathcal D}(\alpha)=\frac{\sum_{\mbox{\scriptsize\boldmath$g$}}P_{e{\mathcal D}}({\mbox{\boldmath$g$}}) \exp\left(-N\alpha(\mbox{\boldmath$g$})\right)}{\sum_{\mbox{\scriptsize\boldmath$g$}}\exp\left(-N\alpha({\mbox{\boldmath$g$}})\right)}.
\end{equation}
Based on a similar result presented in \cite[Lemma 1]{ref Wang11}, the following theorem gives an upper bound on the achievable generalized error performance of a $({\mathcal D}, {\mathcal R}_{\mathcal D})$-decoder.

\begin{theorem}{\label{Theorem3}}
Consider the distributed communication system described in Section \ref{SectionII} with $(K+M)$ transmitters. There exists a decoding algorithm for the $({\mathcal D}, {\mathcal R}_{\mathcal D})$-decoder, such that
\begin{eqnarray}
\label{BoundLemma1}
&& \mbox{GEP}_{\mathcal D}(\alpha) \le \frac{1}{\sum_{\mbox{\scriptsize\boldmath$g$}}\exp\left(-N\alpha({\mbox{\boldmath$g$}})\right)}\Biggl\{  \sum_{\mbox{\scriptsize \boldmath $g$}\in \mathcal{R}_{\mathcal{D}} }  \nonumber \\
&& \sum_{\tiny \begin{array}{c}\mathcal{S}\subset \{1,\cdots,K+M\} \\ \mathcal{D}\setminus\mathcal{S}\ne \emptyset \end{array}} \Biggl[\sum_{\tiny \begin{array}{c} \tilde{\mbox{\scriptsize \boldmath $g$}}\in \mathcal{R}_{\mathcal{D}}, \\ \tilde{\mbox{\scriptsize \boldmath $g$}}_{\mathcal{S}}= \mbox{\scriptsize \boldmath $g$}_{\mathcal{S}} \end{array}} \exp\{-N E_{m\mathcal{D}}(\mathcal{S}, \mbox{\boldmath $g$}, \tilde{\mbox{\boldmath $g$}})\}+     \nonumber \\
&& \left(1+\sum_{\tiny \begin{array}{c} \tilde{\mbox{\scriptsize \boldmath $g$}}\not\in \mathcal{R}_{\mathcal{D}},  \\ \tilde{\mbox{\scriptsize \boldmath $g$}}_{\mathcal{S}}=\mbox{\scriptsize \boldmath $g$}_{\mathcal{S}}\end{array}}1\right) \max_{\tiny \begin{array}{c} \mbox{\scriptsize \boldmath $g$}'\not\in \mathcal{R}_{\mathcal{D}}, \\ \mbox{\scriptsize \boldmath $g$}'_{\mathcal{S}}= \mbox{\scriptsize \boldmath $g$}_{\mathcal{S}} \end{array} } \exp\{-NE_{i\mathcal{D}}(\mathcal{S}, \mbox{\boldmath $g$}, \mbox{\boldmath $g$}') \}\Biggr]\Biggr\} .  \nonumber \\
 \label{SRSBound}
\end{eqnarray}
$E_{m\mathcal{D}}(\mathcal{S}, \mbox{\boldmath $g$}, \tilde{\mbox{\boldmath $g$}})$ and $E_{i\mathcal{D}}(\mathcal{S}, \mbox{\boldmath $g$}, \mbox{\boldmath $g$}')$ in the above equation are given by,
\begin{eqnarray}
\label{EmEiDDcoder}
&& E_{m\mathcal{D}}(\mathcal{S}, \mbox{\boldmath $g$}, \tilde{\mbox{\boldmath $g$}})= \max_{0<\rho \le 1} -\rho \sum_{k\in \mathcal{D}\setminus \mathcal{S}}\tilde{r}_k(\tilde{g}_k)  \nonumber \\
&& + \max_{0<s\le 1} -\log \sum_Y \sum_{\mbox{\scriptsize \boldmath $X$}_{\mathcal{S}}} \prod_{k\in \mathcal{S}\cap\mathcal{D}} P_{X|g_k}(X_k)                    \nonumber \\
&& \left(\sum_{\mbox{\scriptsize \boldmath $X$}_{\mathcal{D}\setminus \mathcal{S}}}\prod_{k \in \mathcal{D}\setminus \mathcal{S}}P_{X|g_k}(X_k)\left[P(Y|\mbox{\boldmath $X$}_{\mathcal{D}}, \mbox{\boldmath $g$}_{\bar{\mathcal{D}}} )e^{-\alpha(\mbox{\scriptsize\boldmath$g$})}\right]^{1-s}\right)     \nonumber \\
&& \left(\sum_{\mbox{\scriptsize \boldmath $X$}_{\mathcal{D}\setminus \mathcal{S}}}\prod_{k \in \mathcal{D}\setminus \mathcal{S}}P_{X|\tilde{g}_k}(X_k)\left[P(Y|\mbox{\boldmath $X$}_{\mathcal{D}}, \tilde{\mbox{\boldmath $g$}}_{\bar{\mathcal{D}}} )e^{-\alpha(\tilde{\mbox{\scriptsize\boldmath$g$}})}\right]^{\frac{s}{\rho}} \right)^{\rho},                       \nonumber \\
&& E_{i\mathcal{D}}(\mathcal{S}, \mbox{\boldmath $g$}, \mbox{\boldmath $g$}') = \max_{0<\rho \le 1} -\rho \sum_{k\in \mathcal{D}\setminus \mathcal{S}}r_k(g_k)     \nonumber \\
&& +\max_{0<s \le 1-\rho} - \log \sum_Y \sum_{\mbox{\scriptsize \boldmath $X$}_{\mathcal{S}}} \prod_{k\in \mathcal{S}\cap\mathcal{D}} P_{X|g_k}(X_k)  \left(\sum_{\mbox{\scriptsize \boldmath $X$}_{\mathcal{D}\setminus \mathcal{S}}}\right.             \nonumber \\
&& \left.\prod_{k \in \mathcal{D}\setminus \mathcal{S}}P_{X|g_k}(X_k)\left[P(Y|\mbox{\boldmath $X$}_{\mathcal{D}}, \mbox{\boldmath $g$}_{\bar{\mathcal{D}}}  )e^{-\alpha(\mbox{\scriptsize\boldmath$g$})} \right]^{\frac{s}{s+\rho}} \right)^{s+\rho}     \nonumber \\
&& \left(\sum_{\mbox{\scriptsize \boldmath $X$}_{\mathcal{D}\setminus \mathcal{S}}}\prod_{k \in \mathcal{D}\setminus \mathcal{S}}P_{X|g'_k}(X_k)P(Y|\mbox{\boldmath $X$}_{\mathcal{D}}, \mbox{\boldmath $g$}'_{\bar{\mathcal{D}}})e^{-\alpha(\mbox{\scriptsize\boldmath$g$}')}\right)^{1-s}, \nonumber \\
\end{eqnarray}
where $r_k(g_k)$, $\tilde{r}_k(\tilde{g}_k)$ are the communications rates corresponding respectively to $g_k$ and $\tilde{g}_k$, and $P(Y|\mbox{\boldmath $X$}_{\mathcal{D}}, \mbox{\boldmath $g$}_{\bar{\mathcal{D}}} )$ is defined as
\begin{equation}
\label{EquivalentCompound}
P(Y|\mbox{\boldmath $X$}_{\mathcal{D}}, \mbox{\boldmath $g$}_{\bar{\mathcal{D}}} )= \sum_{\mbox{\scriptsize \boldmath $X$}_{\bar{\mathcal{D}}}}\prod_{k \in \bar{\mathcal{D}}}P_{X|g_k}(X_k)P_{Y|\mbox{\scriptsize\boldmath$X$}}(Y|\mbox{\boldmath $X$}).
\end{equation}
$\QED$
\end{theorem}

The proof of Theorem \ref{Theorem3} is given in Appendix \ref{ProofTheorem3}.

Let us now come back to the system where the receiver is only interested in decoding the message of transmitter $1$ but can choose to decode the messages of other regular users if necessary. Assume that the receiver is equipped with many $(\mathcal{D}, \mathcal{R}_{\mathcal{D}})$-decoders each corresponding to a transmitter subset $\mathcal{D}\subseteq \{1, \cdots, K\}$ with $1\in \mathcal{D}$ and an operation region $\mathcal{R}_{\mathcal{D}}$. After receiving the channel output symbols, the receiver first carries out all the $(\mathcal{D}, \mathcal{R}_{\mathcal{D}})$-decoding operations. If at least one $(\mathcal{D}, \mathcal{R}_{\mathcal{D}})$-decoder outputs an estimated message and code index pair, and the estimation outputs (i.e., not including the collision reports) of all the $(\mathcal{D}, \mathcal{R}_{\mathcal{D}})$-decoders agree with each other, then the receiver outputs the corresponding estimate $(\hat{w}_1, \hat{g}_1)$ for transmitter $1$. Otherwise, the receiver reports a collision for transmitter $1$.

Let $\mathcal{R}$ be the operation region of the receiver. Since the receiver intends to decode the message of transmitter $1$ if $\mbox{\boldmath $g$}\in \mathcal{R}$, we must have
\begin{equation}
\mathcal{R}\subseteq\bigcup_{\mathcal{D}: \mathcal{D}\subseteq\{1, \cdots, K\}, 1\in \mathcal{D}} \mathcal{R}_{\mathcal{D}},
\label{RegionAssignment}
\end{equation}
On the other hand, for a given $(\mathcal{D}, \mathcal{R}_{\mathcal{D}})$-decoder, since we regard correct message decoding as an expected outcome for $\mbox{\boldmath $g$}\not\in \mathcal{R}_{\mathcal{D}}$, shrinking the operation region of a $(\mathcal{D}, \mathcal{R}_{\mathcal{D}})$-decoder will not hurt its generalized error performance. Consequently, it does not cause any performance degradation to assume that the operation regions of the $(\mathcal{D}, \mathcal{R}_{\mathcal{D}})$-decoders form a partition of $\mathcal{R}$. In other words,
\begin{eqnarray}
&& \mathcal{R}=\bigcup_{\mathcal{D}: \mathcal{D}\subseteq\{1, \cdots, K\}, 1\in \mathcal{D}} \mathcal{R}_{\mathcal{D}}, \qquad \mathcal{R}_{\mathcal{D}'}\cap \mathcal{R}_{\mathcal{D}}=\emptyset, \nonumber \\
&& \quad \forall \mathcal{D}, \mathcal{D}'\subseteq\{1, \cdots, K\}, \mathcal{D}' \ne \mathcal{D}, 1\in \mathcal{D}, \mathcal{D}'.
\label{RegionPartition}
\end{eqnarray}

Based on the above understanding, the following theorem gives an upper bound on the achievable generalized error performance of the receiver.

\begin{theorem}{\label{Theorem4}}
Consider the distributed communication system described in Section \ref{SectionII} with $(K+M)$ transmitters. Assume that the receiver chooses an operation region $\mathcal{R}$. Let $\sigma$ denote a partition of the operation region $\mathcal{R}$ satisfying (\ref{RegionPartition}). There exists a decoding algorithm such that the generalized error performance of the receiver with $\alpha(\mbox{\boldmath $g$})\ge 0$ is upper-bounded by,
\begin{equation}
\mbox{GEP}(\alpha) \le \min_{\sigma} \sum_{\mathcal{D}: \mathcal{D}\subseteq\{1, \cdots, K\}, 1\in \mathcal{D}} \mbox{GEP}_{\mathcal D}(\alpha),
\label{ElementaryGEPBound}
\end{equation}
where $\mbox{GEP}_{\mathcal D}(\alpha)$ is the generalized error performance of the $(\mathcal{D}, \mathcal{R}_{\mathcal{D}})$-decoder, which can be further bounded by (\ref{SRSBound}). $\QED$
\end{theorem}

Theorem \ref{Theorem4} is implied by Theorem \ref{Theorem3}. Note that a simple guideline on how the operation region $\mathcal{R}$ should be chosen is not yet available. We only showed that different choices of $\mathcal{R}$ lead to different error performance bounds.

\section{Collision Detection and Operation Margin}
\label{SectionIV}
Collision detection is a special channel coding task contained in the distributed communication model but not in the coordinated communication model. Because collision report often provides key guidance to medium access control at the data link layer, collision detection details must be carefully investigated to ensure efficient support of upper layer communication adaptation. In the communication error definition specified in Section \ref{SectionII}, given an operation region $\mathcal{R}$, we still regard correct message decoding as an expected communication outcome for $\mbox{\boldmath $g$}\not\in \mathcal{R}$, although the receiver intends to report a collision. Consequently, even if the receiver decodes the message of transmitter $1$, it still cannot conclude with a high probability that the actual code index vector is inside the operation region. In this section, we present extended coding theorems to support reliable collision detection by revising the communication error definition. Further explanations on the necessity of these revisions will be delayed to Section \ref{SectionVI}.

One may think that enforcing collision detection simply means we should exclude correct message decoding from the set of expected outcomes for $\mbox{\boldmath $g$}\not\in \mathcal{R}$, and revise the communication error probability definition to the following
\begin{equation}\label{MC-DecodingErrorDef2}
P_e({\mbox{\boldmath$g$}}) = \left\{\begin{array}{l}\max_{\mbox{\scriptsize\boldmath$w$}}Pr \left\{ (\hat{w}_1,\hat{g}_1) \neq (w_1,g_1)|(\mbox{\boldmath$w$},\mbox{\boldmath$g$}) \right\}, \forall \mbox{\boldmath$g$} \in \mathcal R \\ \max_{\mbox{\scriptsize\boldmath$w$}}1-Pr \left\{ \mbox{``collision''} \right\} \qquad \qquad \forall \mbox{\boldmath$g$} \not\in \mathcal R \end{array} \right.
\end{equation}
Unfortunately, such a simple revision can easily lead to an ill-posed collision detection task, which gives a poor system error performance in the technical result, but may not necessarily reflect the actual performance objective of the system. To illustrate this issue, let us consider a simple example of single user distributed communication over a compound binary symmetric channel. The channel model is given by
\begin{equation}
Y=\left\{\begin{array}{ll}X&\mbox{with probability }1-p\\\bar{X} &\mbox{with probability }p\end{array}\right.,
\end{equation}
where $X, Y\in \{0, 1\}$ are the input and output symbols and $0\le p \le 1$ is the crossover probability that can take four possible values, $p\in \{p_1, p_2, p_3, p_4\}$. Assume that the transmitter has only one coding option, which is a random block code with binary uniform input distribution and communication rate $r$\footnote{Note that we choose to fix the channel code of the transmitter in order to simplify the discussion. The example system can be easily extended to support multiple coding options at the transmitter.}. As explained in Section \ref{SectionII}, impact of the compound channel can be modeled using an interfering user (or a virtual user). Therefore, there are two users in our channel coding model. Transmitter $1$ is the actual regular user with only one coding option, denoted by $g_1=r$. Transmitter $2$ is an interfering user, whose ``code index" parameter $g_2\in \{p_1, p_2, p_3, p_4\}$ is denoted using the compound channel parameter.

Let us first consider the case when $p_1=0.18, p_2= p_3=0.185, p_4=0.19$. In this example, channel realizations $p_2, p_3$ make no statistical difference at the receiver. Assume that $r=0.31$ bit/symbol, which satisfies $1-H(0.19)<r<1-H(0.185)$ where $H()$ is the entropy function. According to Theorem \ref{Theorem1}, region $\tilde{\mathcal R} = \left\{\left[\begin{array}{c}r \\ p_1 \end{array}\right], \left[\begin{array}{c}r \\ p_2 \end{array}\right], \left[\begin{array}{c}r \\ p_3 \end{array}\right] \right\} $ is achievable. However, region $\mathcal R = \left\{\left[\begin{array}{c}r \\ p_1 \end{array}\right], \left[\begin{array}{c}r \\ p_2 \end{array}\right]\right\} \subset \tilde{\mathcal R}$, being a subset of the achievable region $\tilde{\mathcal R}$, is not achievable. This is because choosing $\mathcal R$ as the operation region requires the receiver to guarantee reliable decoding for code index vector $\left[\begin{array}{c}r \\ p_2 \end{array}\right]$ and reliable collision report for $\left[\begin{array}{c}r \\ p_3 \end{array}\right]$. This is not possible since the receiver does not have the capability to distinguish channel realization $p_2$ from $p_3$. When the codeword length is finite, a similar problem exists even when $p_2$ and $p_3$ are slightly different in value. It can be seen that, under error probability definition (\ref{MC-DecodingErrorDef2}), the operation region must be determined carefully to avoid posing a difficult and possibly unecessary detection problem at the receiver. Nevertheless, we will show next that achieving such an objective may not be easy. Suppose that we choose $\tilde{\mathcal R} = \left\{\left[\begin{array}{c}r \\ p_1 \end{array}\right], \left[\begin{array}{c}r \\ p_2 \end{array}\right], \left[\begin{array}{c}r \\ p_3 \end{array}\right]\right\} $ as the operation region, when the codeword length is finite and the actual code index vector is $\left[\begin{array}{c}r \\ p_3 \end{array}\right]$, the receiver has to make sure that channel realization $p_3$ should not be mis-detected as $p_4$. However, such a detection error is difficult to avoid since $p_3$ and $p_4$ are close in value. Similarly, if we choose $\mathcal R = \left\{\left[\begin{array}{c}r \\ p_1 \end{array}\right]\right\} $ as the operation region, and the actual code index vector is $\left[\begin{array}{c}r \\ p_2 \end{array}\right]$, the same problem arises due to the proximity of $p_2$ and $p_1$.

Alternatively, let us still choose $\mathcal R = \left\{\left[\begin{array}{c}r \\ p_1 \end{array}\right]\right\} $ as the operation region but use a slightly revised communication error definition. There are three code index vectors outside the operation region. For $\left[\begin{array}{c}r \\ p_4 \end{array}\right]$, we regard collision report as the only expected outcome. For $\left[\begin{array}{c}r \\ p_2 \end{array}\right]$ and $\left[\begin{array}{c}r \\ p_3 \end{array}\right]$, however, both collision report and correct message decoding are accepted as expected outcomes. We assume that the receiver should decode the message so long as its estimated channel realization is close to $p_1$ and a pre-determined decoding error probability requirement can be met. With such a revision, when the actual code index vector equals $\left[\begin{array}{c}r \\ p_2 \end{array}\right]$ or $\left[\begin{array}{c}r \\ p_3 \end{array}\right]$, so long as the receiver correctly decodes the message of transmitter $1$, the system will not experience a communication error even if the channel realization is not detected precisely. On the other hand, when the actual code index vector equals $\left[\begin{array}{c}r \\ p_1 \end{array}\right]$, messsage decoding is guaranteed unless the receiver erroneously detects the channel as $p_4$. The probability of such a channel detection error is relatively small since $p_1$ and $p_4$ are reasonably apart in value.

According to the above discussion, we revise the channel coding model as follows. Let us assume that, in addition to choosing the operation region $\mathcal{R}$, the receiver chooses another region $\widehat{\mathcal{R}}$, termed the ``operation margin", that is non-overlapping with the operation region, i.e., $\mathcal{R}\cap \widehat{\mathcal{R}}=\emptyset$. The receiver intends to decode the message of transmitter $1$ for $\mbox{\boldmath $g$}\in \mathcal{R}$, and to report a collision for $\mbox{\boldmath $g$}\not\in \mathcal{R}\cup \widehat{\mathcal{R}}$. While for $\mbox{\boldmath $g$}\in \widehat{\mathcal{R}}$, both correct message decoding and collision report are accepted as expected outcomes. In each time slot, upon receiving the channel output symbols $\mbox{\boldmath$y$}$, the receiver finds within the operation region $\mathcal{R}$ an estimate of the code index vector, denoted by $\hat{\mbox{\boldmath$g$}}\in \mathcal{R}$. The receiver outputs the estimated message and code index of transmitter $1$, denoted by $(\hat{w}_1, \hat{g}_1)$, if a pre-determined decoding and code index estimation error probability requirement can be met. Otherwise, the receiver reports a collision for transmitter $1$. The purpose of introducing the operation margin is to create a buffer zone between the operation region $\mathcal{R}$, where correct message decoding should be enforced, and the region $\overline{\mathcal{R}\cup \widehat{\mathcal{R}}}$, where collision report should be enforced. Providing the receiver with the option of moving some of the code index vectors into the operation margin $\widehat{\mathcal{R}}$ can help to avoid the ill-posed collision detection problem illustrated above, while still allows the requirement of estimating certain communication parameters in the collision detection task. Note that the revised system model is an extension to the one considered in Sections \ref{SectionII} and \ref{SectionIII} since the latter can be viewed as choosing $\widehat{\mathcal{R}}$ as the compliment of $\mathcal{R}$, i.e., $\widehat{\mathcal{R}}=\bar{\mathcal{R}}$. Also note that, similar to Section \ref{SectionIII}, a simple guideline on how the operation margin $\widehat{\mathcal{R}}$ should be chosen is not yet available. We will only show that different choices of $\mathcal{R}$ and $\widehat{\mathcal{R}}$ lead to different error performance bounds.

Given $\mathcal{R}$ and $\widehat{\mathcal{R}}$, communication error probability as a function of $\mbox{\boldmath$g$}$ is given by
\begin{equation}\label{MC-DecodingErrorDef3}
P_e({\mbox{\boldmath$g$}}) = \left\{\begin{array}{l}\max_{\mbox{\scriptsize\boldmath$w$}}Pr \left\{ (\hat{w}_1,\hat{g}_1) \neq (w_1,g_1)|(\mbox{\boldmath$w$},\mbox{\boldmath$g$}) \right\}, \forall \mbox{\boldmath$g$} \in \mathcal R \\ \max_{\mbox{\scriptsize\boldmath$w$}}1-Pr \left\{ \mbox{``collision''} \mbox{ or } \right.       \\ \quad       \left. (\hat{w}_1,\hat{g}_1) =(w_1,g_1)|(\mbox{\boldmath$w$},\mbox{\boldmath$g$}) \right\},  \qquad \forall \mbox{\boldmath$g$} \in \widehat{\mathcal{R}} \\ \max_{\mbox{\scriptsize\boldmath$w$}}1-Pr \left\{ \mbox{``collision''} \right\}  \qquad \forall \mbox{\boldmath$g$} \not\in \mathcal{R}\cup \widehat{\mathcal{R}} \end{array} \right.
\end{equation}
Define the system error probability and the generalized error performance measure as in (\ref{SystemError}) and (\ref{GenerlizedError}). Let us fix the coding parameters that are not functions of the codeword length. We say an operation region and operation margin pair $(\mathcal{R}, \widehat{\mathcal{R}})$ is achievable if there exits a set of decoding algorithms whose system error probability converges to zero as the codeword length is taken to infinity. The following theorem is an extended version of Theorem \ref{Theorem1} for the revised system model.
\begin{theorem}\label{Theorem5}
Consider the distributed communication system described in Section \ref{SectionIV} with $(K+M)$ transmitters. Let the operation region $\mathcal{R}$ be given by (\ref{MutualInformationInequality1}). Any operation region and operation margin pair $(\mathcal{R}, \widehat{\mathcal{R}})$ with an arbitrary choice of $\widehat{\mathcal{R}}$ is achievable. $\QED$
\end{theorem}

Theorem \ref{Theorem5} can be proven by following the same proof of Theorem \ref{Theorem1}.

Similar to Theorem \ref{Theorem2}, the following theorem is implied directly by the achievable region definition.

\begin{theorem}\label{Theorem6}
For the distributed communication system considered in Theorem \ref{Theorem5}, if an operation region and operation margin pair $(\mathcal{R}, \widehat{\mathcal{R}})$  is achievable, then any other operation region and operation margin pair $(\mathcal{R}_1, \widehat{\mathcal{R}}_1)$ that satisfies $\mathcal{R}_1\subseteq \mathcal{R}$ and $\mathcal{R}_1 \cup \widehat{\mathcal{R}}_1 \supseteq \mathcal{R} \cup \widehat{\mathcal{R}}$ is also achievable. $\QED$
\end{theorem}

When the codeword length is finite, given the operation region $\mathcal{R}$ and the operation margin $\widehat{\mathcal{R}}$, we again decompose the decoder into a set of ``$({\mathcal D}, {\mathcal R}_{\mathcal D})$-decoders" for all ${\mathcal D}\subseteq \{1, \cdots, K\}$ with $1\in \mathcal{D}$. For each $({\mathcal D}, {\mathcal R}_{\mathcal D})$-decoder, we denote its operation region by ${\mathcal R}_{\mathcal D}$ and set its operation margin as $\widehat{\mathcal{R}}_{\mathcal D}=(\mathcal{R}\cup\widehat{\mathcal{R}})\setminus{\mathcal R}_{\mathcal D}$. Decoding procedure of the $({\mathcal D}, {\mathcal R}_{\mathcal D})$-decoder is the same as described in Section \ref{SectionIII}, with the communication error probability being defined as,
\begin{eqnarray}\label{MC-DecodingErrorDefDRD2}
&& P_{e{\mathcal D}}({\mbox{\boldmath$g$}})=    \nonumber \\
&& \quad \left\{ \begin{array}{l}\max_{\mbox{\scriptsize\boldmath$w$}}Pr \left\{ (\hat{\mbox{\boldmath$w$}}_{\mathcal D}, \hat{\mbox{\boldmath$g$}}_{\mathcal D}) \neq (\mbox{\boldmath$w$}_{\mathcal D}, \mbox{\boldmath$g$}_{\mathcal D})|(\mbox{\boldmath$w$},\mbox{\boldmath$g$}) \right\},       \\ \qquad \qquad \qquad \qquad \qquad \qquad \qquad        \forall \mbox{\boldmath$g$} \in \mathcal R_{\mathcal D} \\ \max_{\mbox{\scriptsize\boldmath$w$}}1-Pr \left\{ \mbox{``collision''} \mbox{ or } \right.     \\ \quad      \left. (\hat{\mbox{\boldmath$w$}}_{\mathcal D}, \hat{\mbox{\boldmath$g$}}_{\mathcal D}) = (\mbox{\boldmath$w$}_{\mathcal D}, \mbox{\boldmath$g$}_{\mathcal D}) |(\mbox{\boldmath$w$},\mbox{\boldmath$g$}) \right\} \quad \forall \mbox{\boldmath$g$} \in \widehat{\mathcal{R}}_{\mathcal D} \\ \max_{\mbox{\scriptsize\boldmath$w$}}1-Pr \left\{ \mbox{``collision''}\right\} \qquad \forall \mbox{\boldmath$g$} \not\in  \mathcal R_{\mathcal D}\cup\widehat{\mathcal{R}}_{\mathcal D} \end{array} \right. .
\end{eqnarray}
The following theorem gives an upper bound on the achievable generalized error performance, defined in (\ref{GEPDRD}), of a $({\mathcal D}, {\mathcal R}_{\mathcal D})$-decoder.

\begin{theorem}{\label{Theorem7}}
Consider the distributed communication system described in Section \ref{SectionIV} with $(K+M)$ transmitters and a $({\mathcal D}, {\mathcal R}_{\mathcal D})$-decoder whose operation region and operation margin are denoted by ${\mathcal R}_{\mathcal D}$ and $\widehat{\mathcal{R}}_{\mathcal D}$ respectively. There exists a decoding algorithm whose generalized error performance satisfies the following bound.
\begin{eqnarray}
\label{BoundLemma2}
&& \mbox{GEP}_{\mathcal D}(\alpha) \le \frac{1}{\sum_{\mbox{\scriptsize\boldmath$g$}}\exp\left(-N\alpha({\mbox{\boldmath$g$}})\right)}\Biggl\{\sum_{\mbox{\scriptsize \boldmath $g$}\in \mathcal{R}_{\mathcal{D}} } \nonumber \\
&& \sum_{\tiny \begin{array}{c}\mathcal{S}\subset \{1,\cdots,K+M\} \\ \mathcal{D}\setminus\mathcal{S}\ne \emptyset \end{array}} \Biggl[\sum_{\tiny \begin{array}{c} \tilde{\mbox{\scriptsize \boldmath $g$}}\in \mathcal{R}_{\mathcal{D}}, \\ \tilde{\mbox{\scriptsize \boldmath $g$}}_{\mathcal{S}}= \mbox{\scriptsize \boldmath $g$}_{\mathcal{S}} \end{array}} \exp\{-N E_{m\mathcal{D}}(\mathcal{S}, \mbox{\boldmath $g$}, \tilde{\mbox{\boldmath $g$}})\}+     \nonumber \\
&& \left(1+\sum_{\tiny \begin{array}{c} \tilde{\mbox{\scriptsize \boldmath $g$}}\not\in \mathcal{R}_{\mathcal{D}},  \\ \tilde{\mbox{\scriptsize \boldmath $g$}}_{\mathcal{S}}=\mbox{\scriptsize \boldmath $g$}_{\mathcal{S}}\end{array}}1\right) \max_{\tiny \begin{array}{c} \mbox{\scriptsize \boldmath $g$}'\not\in \mathcal{R}_{\mathcal{D}}, \\ \mbox{\scriptsize \boldmath $g$}'_{\mathcal{S}}= \mbox{\scriptsize \boldmath $g$}_{\mathcal{S}} \end{array} } \exp\{-NE_{i\mathcal{D}}(\mathcal{S}, \mbox{\boldmath $g$}, \mbox{\boldmath $g$}') \}\Biggr] \nonumber \\
&&  + \sum_{\mbox{\scriptsize \boldmath $g$}\in \mathcal{R}_{\mathcal{D}} }\sum_{\tiny \begin{array}{c}\mathcal{S}\subset \{1,\cdots,K+M\} \\ \mathcal{D}\setminus\mathcal{S}= \emptyset \end{array}} \Biggl[ \left(1+\sum_{\tiny \begin{array}{c} \tilde{\mbox{\scriptsize \boldmath $g$}}\not\in \mathcal{R}_{\mathcal{D}}\cup \widehat{\mathcal{R}}_{\mathcal D},  \\ \tilde{\mbox{\scriptsize \boldmath $g$}}_{\mathcal{S}}=\mbox{\scriptsize \boldmath $g$}_{\mathcal{S}}\end{array}}1\right)     \nonumber \\
&&  \times\max_{\tiny \begin{array}{c} \mbox{\scriptsize \boldmath $g$}'\not\in \mathcal{R}_{\mathcal{D}}\cup \widehat{\mathcal{R}}_{\mathcal D} , \\ \mbox{\scriptsize \boldmath $g$}'_{\mathcal{S}}= \mbox{\scriptsize \boldmath $g$}_{\mathcal{S}} \end{array} } \exp\{-NE_{i\mathcal{D}}(\mathcal{S}, \mbox{\boldmath $g$}, \mbox{\boldmath $g$}') \} \Biggr] \Biggr\}.
 \label{SRSBound2}
\end{eqnarray}
$E_{m\mathcal{D}}(\mathcal{S}, \mbox{\boldmath $g$}, \tilde{\mbox{\boldmath $g$}})$ and $E_{i\mathcal{D}}(\mathcal{S}, \mbox{\boldmath $g$}, \mbox{\boldmath $g$}')$ in the above equation are given by (\ref{EmEiDDcoder}). $P(Y|\mbox{\boldmath $X$}_{\mathcal{D}}, \mbox{\boldmath $g$}_{\bar{\mathcal{D}}} )$ is defined as in (\ref{EquivalentCompound}).
$\QED$
\end{theorem}

The proof of Theorem \ref{Theorem7} is given in Appendix \ref{ProofTheorem7}.

With Theorem \ref{Theorem7}, a performance bound at the receiver can be derived in a way similar to that in Section \ref{SectionIII}. Let the operation region and the operation margin of the receiver be given by $\mathcal{R}$ and $\widehat{\mathcal{R}}$. Assume that the receiver is equipped with many $(\mathcal{D}, \mathcal{R}_{\mathcal{D}})$-decoders each corresponding to a transmitter subset $\mathcal{D}\subseteq \{1, \cdots, K\}$ with $1\in \mathcal{D}$. Given the operation region $\mathcal{R}_{\mathcal{D}}$ of an $(\mathcal{D}, \mathcal{R}_{\mathcal{D}})$-decoder, we set its operation margin at $\widehat{\mathcal{R}}_{\mathcal D}=(\mathcal{R}\cup\widehat{\mathcal{R}})\setminus{\mathcal R}_{\mathcal D}$. By following the same decoding algorithm and the same discussion as presented in Section \ref{SectionIII}, we can see that it does not cause any performance degradation to let the operation regions of the $(\mathcal{D}, \mathcal{R}_{\mathcal{D}})$-decoders form a partition of $\mathcal{R}$. Consequently, an upper bound on the achievable generalized error performance of the receiver can be obtained, as stated in the following theorem.

\begin{theorem}{\label{Theorem8}}
Consider the distributed communication system described in Section \ref{SectionIV} with $(K+M)$ transmitters. Assume that the receiver chooses an operation region $\mathcal{R}$ and an operation margin $\widehat{\mathcal{R}}$ with $\mathcal{R}\cap\widehat{\mathcal{R}}=\emptyset$. Let $\sigma$ be a partition of the operation region $\mathcal{R}$ satisfying (\ref{RegionPartition}). There exists a decoding algorithm such that the generalized error performance of the receiver with $\alpha(\mbox{\boldmath $g$})\ge 0$ is upper-bounded by (\ref{ElementaryGEPBound}) with $\mbox{GEP}_{\mathcal D}(\alpha)$ being further bounded by (\ref{SRSBound2}). $\QED$
\end{theorem}

Theorem \ref{Theorem8} is implied by Theorem \ref{Theorem7}.

Note that the generalized error performance bounds given in Theorems \ref{Theorem4} and \ref{Theorem8} are implicit since the optimal partition $\sigma$ that maximizes the right hand side of (\ref{ElementaryGEPBound}) is not specified. To find the optimal partition, one needs to compute every single term on the right hand side of (\ref{ElementaryGEPBound}), (\ref{SRSBound}) and (\ref{SRSBound2}) for all code index vectors and all transmitter subsets. Because each term in the definitions of $E_{m\mathcal{D}}(\mathcal{S}, \mbox{\boldmath $g$}, \tilde{\mbox{\boldmath $g$}})$ and $E_{i\mathcal{D}}(\mathcal{S}, \mbox{\boldmath $g$}, \mbox{\boldmath $g$}')$ involves the combinations of one transmitter subset and two code index vectors, the computational complexity of finding the optimal partition is therefore in the order of $O\left(2^K \left(\prod_{k=1}^{K+M}|{\mathcal G}_k|\right)^2\right)$.

\section{Coding Complexity and Channel Code Detection}
\label{SectionV}

Computational complexity is one of the key factors that must be considered in the development of a channel coding scheme. Because a distributed communication system often deals with packets (and therefore codewords) that are short in length, complexity problem in the new coding model is quite different from the classical ones. In this section, we will discuss an important complexity aspect of the coding model. Note that, a thorough investigation on low complexity coding in distributed communication, although being important, is beyond the scope of this paper.

According to the decoding algorithms presented in Appendices \ref{ProofTheorem3} and \ref{ProofTheorem7}, upon receiving the channel output symbols, the receiver needs to compute the likelihood of all codewords corresponding to all code index vectors in the operation region. The complexity of such a decoding algorithm is in the order of $O\left(\sum_{\mbox{\scriptsize \boldmath $g$}\in \mathcal{R}}\exp\left(N\left(\sum_{k=1}^K r_k(g_k)\right)\right)\right)$. It is important to note that the number of code index vectors in the operation region can be excessive. First, a receiver in a distributed communication system does not necessarily know which transmitters will be active in the area. By taking potential transmitters into decoding consideration, the number of transmitters in the channel coding model can be much larger than the number of active transmitters. Second, channel coding in a distributed communication system equips a transmitter with multiple coding options. If the system should be prepared for a wide range of communication environments, then the set of coding options of each user can have a large cardinality.

A simple way to avoid calculating the likelihood of too many codewords in channel decoding is to first let the receiver detect the code index vector using the distribution information of channel input and output symbols. The receiver can then process only the codewords corresponding to the detected code indices. Note that, without message decoding, the receiver may not have the capability to estimate the code index vector precisely. For example, in a system with homogeneous transmitters, a receiver may not be able to tell the identities of the active transmitters based only on the distribution information of the channel input and output symbols. Therefore, a reasonable approach is to carefully partition the space of code index vectors into several regions and to detect the region where the code index vector belongs. Such a detection outcome can still help to significantly reduce the number of codewords that should be further processed by the receiver. In addition to complexity reduction, code index vector detection is also useful for other system functions such as communication adaptation at the data link layer.

Let us assume that the receiver partition the space of code index vectors into $L$ regions, denoted by $C_1, \cdots, C_L$. Let $\mbox{\boldmath $g$}\in C$ (with $C\in \{ C_1, \cdots, C_L\}$) be the actual code index vector. Given the channel output $\mbox{\boldmath $y$}$ and the distribution information of the codebooks, the receiver wants to detect the region to which the code index vector belongs. Given $\alpha(\mbox{\boldmath $g$})\ge 0$, if code index vector $\mbox{\boldmath $g$}$ is chosen with a prior probability of $\frac{\exp\left(-N\alpha(\mbox{\scriptsize \boldmath$g$})\right)}{\sum_{\mbox{\scriptsize\boldmath$g$}}\exp\left(-N\alpha({\mbox{\scriptsize\boldmath$g$}})\right)}$, then the optimal estimate of the code index vector $\hat{\mbox{\boldmath $g$}}$ is given by.
\begin{equation}
\hat{\mbox{\boldmath $g$}}=\mathop{\mbox{argmax}}_{\tilde{\mbox{\scriptsize \boldmath $g$}}} P(\mbox{\boldmath $y$}| \tilde{\mbox{\boldmath $g$}})e^{-N\alpha(\tilde{\mbox{\scriptsize\boldmath $g$}} )}
\label{MAPgestimate}
\end{equation}
We say that the region detection is successful if $\hat{\mbox{\boldmath $g$}}\in C$. The following theorem gives an upper bound to the detection error probability as a function of $\mbox{\boldmath $g$}$.

\begin{theorem}{\label{Theorem9}}
Consider the distributed communication system described in Section \ref{SectionIV} with $(K+M)$ transmitters and with the code index region detection described above. Let $\mbox{\boldmath $g$}$ be the actual code index vector, which belongs to region $C$. The probability that $\hat{\mbox{\boldmath $g$}}$ given by (\ref{MAPgestimate}) does not belong to $C$ satisfies the following bound
\begin{equation}\label{CommParameterDetectionBound}
Pr\{\hat{\mbox{\boldmath $g$}}\not\in C|\mbox{\boldmath $g$}\}e^{-N\alpha(\mbox{\scriptsize\boldmath $g$})}\le  \sum_{\tilde{\mbox{\scriptsize \boldmath $g$}}\not\in C}\exp\left(-NE_c(\mbox{\boldmath $g$}, \tilde{\mbox{\boldmath $g$}})\right),
\end{equation}
with
\begin{eqnarray}
&& E_c(\mbox{\boldmath $g$}, \tilde{\mbox{\boldmath $g$}})=\max_{0<s\le 1}-\log\Biggl(\sum_Y \left[P(Y|\mbox{\boldmath $g$})e^{-\alpha(\mbox{\scriptsize\boldmath $g$})}\right]^{s} \nonumber \\
&& \quad \times \left[P(Y|\tilde{\mbox{\boldmath $g$}})e^{-\alpha(\tilde{\mbox{\scriptsize\boldmath $g$}})}\right]^{(1-s)}\Biggr).
\end{eqnarray}
$\QED$
\end{theorem}

The proof of Theorem \ref{Theorem9} is given in Appendix \ref{ProofTheorem9}.

Assume that a receiver first detect the region to which the code index vector belongs, and then search decoding output among codewords corresponding to code index vectors inside the detected region. Performance bound of such a receiver can be easily derived by combining the results of Theorems \ref{Theorem3}, \ref{Theorem7} and \ref{Theorem9}. Computational complexity of the decoding algorithm is reduced to the order of $O\left(\max_{i\in\{1, \cdots, L\}}\sum_{\mbox{\scriptsize \boldmath $g$}\in C_i\cap\mathcal{R}}\exp\left(N\left(\sum_{k=1}^K r_k(g_k)\right)\right)\right)$. Note that, the complexity reduction due to code index detection may not appear to be significant in the above expression. However, such a picture can change easily if the complexity scaling law in the codeword length can be reduced from exponential to polynomial.

Similar to the collision detection problem, one should note that the code index detection problem can also become ill-posed, for a reason similar to the one discussed in Section \ref{SectionIV}. Take the example with four similar compound channel gains discussed in Section \ref{SectionIV}, it is easy to see that any none trivial partitioning of the code index space will lead to a poor code index detection performance. The solution to such an issue is to follow the idea of ``operation margin" introduction and, for every code index region, to mark some other regions as its detection margin. In stead of distinguishing code index vectors between different regions, one can relax the detection problem and only require the receiver to distinguish code index vectors inside a region from those outside the region and the detection margin. Further discussion on this issue and the corresponding performance bound derivation are quite straightforward and are therefore skipped in the paper.

\section{Further Discussions}
\label{SectionVI}
The new channel coding framework, originally proposed in \cite{ref Luo12}\cite{ref Wang12} and generalized in this paper, allows a transmitter in a distributed communication system to choose its channel code without sharing such a decision with other transmitters or with the receiver. A potential impact of the new coding theorems, which is also a key motivation of their development, is that they can be exploited to enhance the interface between the physical layer and the data link layer in the architecture of wireless networks. More specifically, by forwarding certain freedom of channel code determination to the data link layer, a data link layer user can be equipped with a handful of transmission options corresponding to different values of communication parameters such as power, rate and antenna beam. This is opposed to the binary transmitting/idling options widely assumed at the data link layer in the current network architecture. Depending on the transmission choices of the users, outcomes at the link layer receivers can be analyzed explicitly using physical layer channel properties. Consequently, advanced communication adaptation approaches such as rate and power adaptation for efficient medium access control can be considered at the link layer. Because collision report and coding parameter estimation provide important guidance to link layer communication adaptation, whether and how these detection tasks should be enforced at a physical layer receiver must be carefully investigated in channel coding problem formulation. This is the concern behind different communication error definitions given in equations (\ref{MC-DecodingErrorDef}), (\ref{MC-DecodingErrorDefDRD}), (\ref{MC-DecodingErrorDef2}), (\ref{MC-DecodingErrorDef3}) and (\ref{MC-DecodingErrorDefDRD2}).

Error probability bounds presented in this paper are derived based on a random coding scheme using techniques revised from those of \cite{ref Gallager65}. The non-asymptotic bounds are valid for any codeword length. They are also easy to evaluate numerically. However, it is well known that, in a single-user communication system, random coding does not achieve the best error exponent at low information rates. Error exponents obtained using a random coding scheme can be less insightful than those obtained using constant composition codes \cite{ref Csiszar11}. In fact, it was shown in \cite{ref Farkas14} that, with the help of constant composition codes at the transmitters and maximum mutual information decoder at the receiver, decoding error exponents obtained in \cite{ref Wang12} can be further improved. Expressions of the error exponents can also be simplified due to an interesting decoupling property \cite{ref Farkas14}.

Coding results presented in this paper and in \cite{ref Luo12}\cite{ref Wang12} provided basic understandings about channel coding in distributed communication systems. They also lead to many new research problems that need to be investigated further. At the physical layer, distributed communication requires a transmitter to prepare a code library before knowing the communication environment. How should such a code library be chosen is a problem that has never be addressed before. At the data link layer, when each transmitter is equipped with multiple transmission options, in the case of packet collision, a transmitter has the choice of switching to other transmission options as opposed to simply ``backing-off" by reducing its transmission probability. How to exploit multiple transmission options to improve the efficiency of medium access control is a problem that received little attention before. An early investigation on link layer communication adaptation with an enhanced physical-link layer interface was reported only recently in \cite{ref Tang14}.

\section{Conclusion}
\label{SectionVII}
We presented a generalized channel coding model for distributed wireless communication and derived its performance limits and tradeoff bounds. Key ideas behind these results are the combination of message decoding and collision detection, and the extension of communication error definition beyond the classical meaning of decoding failure. Our results demonstrated that, by matching the communication error definition with that of the unexpected system outcome, classical channel coding theorems can potentially be extended to a wide range of communication modules. It is our hope that the results and the analytical framework presented in this paper can serve as a bridge that eventually leads to rigorous understandings about channel coding and its impact in distributed network systems.


\appendix
\subsection{Proof of Theorem \ref{Theorem3}}
\label{ProofTheorem3}
\begin{proof}
Given a user subset $\mathcal S \subset \{1, \cdots, K+M\}$ with $\mathcal D \setminus \mathcal S \ne \emptyset$, we first define in the following a notation $(\mbox{\boldmath$w$}_{\mathcal D},\mbox{\boldmath$g$}) \stackrel{\mathcal S}{=} (\tilde{\mbox{\boldmath$w$}}_{\mathcal D},\tilde{\mbox{\boldmath$g$}})$ that will significantly simplify the expressions in the proof.
\begin{eqnarray}\label{WsGsEquality}
&& (\mbox{\boldmath$w$}_{\mathcal D},\mbox{\boldmath$g$}) \stackrel{\mathcal S}{=} (\tilde{\mbox{\boldmath$w$}}_{\mathcal D},\tilde{\mbox{\boldmath$g$}}) :      \nonumber \\
&& \quad           (\mbox{\boldmath$w$}_{\mathcal{S}\cap \mathcal{D}},\mbox{\boldmath$g$}_{\mathcal{S}\cap \mathcal{D}}) = (\tilde{\mbox{\boldmath$w$}}_{\mathcal{S}\cap \mathcal{D}}, \tilde{\mbox{\boldmath$g$}}_{\mathcal{S}\cap \mathcal{D}}), \mbox{\boldmath$g$}_{\mathcal{S}\cap \bar{\mathcal{D}}}=\tilde{\mbox{\boldmath$g$}}_{\mathcal{S}\cap \bar{\mathcal{D}}}, \nonumber \\
&& \quad (w_k,g_k)\neq(\tilde{w}_{k},\tilde{g}_k), \forall k \in \mathcal D \setminus \mathcal S, g_k \neq \tilde{g}_k, \forall k \in \bar{\mathcal D} \setminus \mathcal S . \nonumber \\
\end{eqnarray}

We assume that the following decoding algorithm is used at the receiver. Given the received channel output symbols $\mbox{\boldmath$y$}$, the receiver estimates the code index vector $\mbox{\boldmath$g$}$. The receiver outputs the messages and code index estimates for users in $\mathcal D$, denoted jointly by $(\mbox{\boldmath$w$}_{\mathcal D}, \mbox{\boldmath$g$}_{\mathcal D})$, if $\mbox{\boldmath$g$} \in \mathcal R_{\mathcal D}$ and the following condition is satisfied for all user subsets $\mathcal{S} \subset \{1,\cdots,K+M\}$ with $\mathcal D \setminus \mathcal S \ne \emptyset$.
\begin{eqnarray}
\label{CompoundCriterionAppendix}
&& -\frac{1}{N}\log P(\mbox{\boldmath $y$}|\mbox{\boldmath $x$}_{(\mbox{\scriptsize\boldmath$w$}_{\mathcal D},\mbox{\scriptsize\boldmath$g$}_{\mathcal D})}, \mbox{\boldmath$g$}_{\bar{\mathcal D}})+\alpha(\mbox{\boldmath$g$})      \nonumber \\
&& \quad < -\frac{1}{N}\log P(\mbox{\boldmath $y$}|\mbox{\boldmath $x$}_{(\tilde{\mbox{\scriptsize\boldmath$w$}}_{\mathcal D}, \tilde{\mbox{\scriptsize\boldmath$g$}}_{\mathcal D})},\tilde{\mbox{\boldmath$g$}}_{\bar{\mathcal D}})+\alpha(\tilde{\mbox{\boldmath$g$}}) , \nonumber \\
&&\mbox{ for all } (\tilde{\mbox{\boldmath$w$}}_{\mathcal D},\tilde{\mbox{\boldmath$g$}}), (\tilde{\mbox{\boldmath$w$}}_{\mathcal D},\tilde{\mbox{\boldmath$g$}}) \stackrel{\mathcal S}{=} (\mbox{\boldmath$w$}_{\mathcal D},\mbox{\boldmath$g$})      \nonumber \\
&& \quad \mbox{ and }(\tilde{\mbox{\boldmath$w$}}_{\mathcal D}, \tilde{\mbox{\boldmath$g$}}), (\mbox{\boldmath$w$}_{\mathcal D}, \mbox{\boldmath$g$}) \in \mathcal{R}_{(\mathcal{S},\mbox{\scriptsize \boldmath $y$})}, \mbox{ with} \nonumber\\
&& \mathcal{R}_{(\mathcal{S},\mbox{\scriptsize \boldmath $y$})}=\Bigl\{\left.(\tilde{\mbox{\boldmath$w$}}_{\mathcal D}, \tilde{\mbox{\boldmath$g$}})\right| \tilde{\mbox{\boldmath $g$}}\in \mathcal{R}_{\mathcal D}, \nonumber \\
&& \quad -\frac{1}{N}\log P(\mbox{\boldmath $y$}|\mbox{\boldmath $x$}_{(\tilde{\mbox{\scriptsize\boldmath$w$}}_{\mathcal D}, \tilde{\mbox{\scriptsize\boldmath$g$}}_{\mathcal D})}, \tilde{\mbox{\boldmath$g$}}_{\bar{\mathcal D}})+\alpha(\tilde{\mbox{\boldmath$g$}}) \nonumber \\
&& \quad \left.  < \tau_{(\tilde{\mbox{\scriptsize\boldmath$g$}},\mathcal{S})}(\mbox{\boldmath$x$}_{{\mathcal S} \cap {\mathcal D} }, \mbox{\boldmath$y$}) \right\},
\end{eqnarray}
$\tau_{(\tilde{\mbox{\scriptsize\boldmath$g$}},\mathcal{S})}(\cdot)$ in the above equation is a pre-determined typicality threshold function of $(\mbox{\boldmath$x$}_{{\mathcal S} \cap {\mathcal D} }, \mbox{\boldmath$y$})$, associated with code index vector $\tilde{\mbox{\boldmath$g$}}$ and user subset $\mathcal{S}$. The determination of $\tau_{(\tilde{\mbox{\scriptsize\boldmath$g$}},\mathcal{S})}(\cdot)$ will be discussed in Step IV of the proof. If there is no codeword satisfying (\ref{CompoundCriterionAppendix}), the receiver reports a collision. In other words, for a given $\mathcal{S}$, the receiver searches for the subset of codewords whose weighted likelihood values are larger than the corresponding typicality threshold. If the subset is not empty, the receiver outputs the codeword with the maximum posterior probability (or weighted likelihood) as the estimate, for this given $\mathcal{S}$. If the receiver outputs an estimate (i.e., not a collision) for at least one $\mathcal{S}$ and the estimates for all $\mathcal{S} \subset \{1,\cdots,K+M\}$ with $\mathcal D \setminus \mathcal S \ne \emptyset$ agree with each other, the receiver regards this estimate as the decoding decision and outputs the corresponding decoded message and code index vector pair. Otherwise, the receiver reports a collision. Note that in (\ref{CompoundCriterionAppendix}), for given $\mathcal{S}$ and ($\mbox{\boldmath$w$}_{\mathcal D},\mbox{\boldmath$g$}$), we only compare the weighted likelihood of codeword vector $\mbox{\boldmath$x$}_{(\mbox{\scriptsize\boldmath$w$}_{\mathcal D},\mbox{\scriptsize\boldmath$g$}_{\mathcal D})}$ with other codeword vectors satisfying $(\tilde{\mbox{\boldmath$w$}}_{\mathcal D},\tilde{\mbox{\boldmath$g$}}) \stackrel{\mathcal S}{=} (\mbox{\boldmath$w$}_{\mathcal D},\mbox{\boldmath$g$})$. We will first analyze the error performance for each user subset $\mathcal{S}$ and then derive the overall error performance by taking the union over all $\mathcal{S}$.

Given a user subset $\mathcal{S}$ with $\mathcal D \setminus \mathcal S \ne \emptyset$, we define the following probability terms.

First, assume that $\mbox{\boldmath$w$}_{\mathcal D}$ is the transmitted message vector for users in $\mathcal D$, and $\mbox{\boldmath$g$}$ is the actual code index vector with $\mbox{\boldmath$g$} \in \mathcal{R}_{\mathcal D}$. Let $P_{t[\mbox{\scriptsize\boldmath$g$},\mathcal{S}]}$ be the probability that the weighted likelihood of the transmitted codeword vector is no larger than the corresponding typicality threshold,
\begin{eqnarray}
\label{ProofPtCC}
&& P_{t[\mbox{\scriptsize\boldmath$g$},\mathcal{S}]} = \nonumber \\
&& Pr \left\{ P(\mbox{\boldmath $y$}|\mbox{\boldmath $x$}_{(\mbox{\scriptsize\boldmath$w$}_{\mathcal D},\mbox{\scriptsize\boldmath$g$}_{\mathcal D})} ,\mbox{\boldmath$g$}_{\bar{\mathcal D}}) \le e^{-N[\tau_{(\mbox{\tiny\boldmath$g$},\mathcal{S})}(\mbox{\scriptsize\boldmath$x$}_{{\mathcal S} \cap {\mathcal D} }, \mbox{\scriptsize\boldmath$y$})-\alpha(\mbox{\scriptsize\boldmath$g$})]}\right\}. \nonumber \\
\end{eqnarray}
Define $P_{m[\mbox{\scriptsize\boldmath$g$},\tilde{\mbox{\scriptsize\boldmath$g$}},\mathcal{S}]}$ as the probability that the posterior probability (or weighted likelihood) of the transmitted codeword vector is no larger than that of another codeword vector $(\tilde{\mbox{\boldmath$w$}}_{\mathcal D},\tilde{\mbox{\boldmath$g$}})$ with $(\tilde{\mbox{\boldmath$w$}}_{\mathcal D},\tilde{\mbox{\boldmath$g$}}) \stackrel{\mathcal S}{=} (\mbox{\boldmath$w$}_{\mathcal D},\mbox{\boldmath$g$})$ and $\tilde{\mbox{\boldmath$g$}} \in \mathcal{R}_{\mathcal D}$,
\begin{eqnarray}
\label{ProofPmMC}
&& P_{m[\mbox{\scriptsize\boldmath$g$},\tilde{\mbox{\scriptsize\boldmath$g$}},\mathcal{S}]} =  Pr \left\{ P(\mbox{\boldmath $y$}|\mbox{\boldmath $x$}_{(\mbox{\scriptsize\boldmath$w$}_{\mathcal D},\mbox{\scriptsize\boldmath$g$}_{\mathcal D})}, \mbox{\boldmath$g$}_{\bar{\mathcal D}})e^{-N\alpha(\mbox{\scriptsize\boldmath$g$})} \right.      \nonumber \\
&& \quad \left. \le P(\mbox{\boldmath $y$}|\mbox{\boldmath $x$}_{(\tilde{\mbox{\scriptsize\boldmath$w$}}_{\mathcal D}, \tilde{\mbox{\scriptsize\boldmath$g$}}_{\mathcal D})},\tilde{\mbox{\boldmath$g$}}_{\bar{\mathcal D}})e^{-N\alpha(\tilde{\mbox{\scriptsize\boldmath$g$}})}\right\}\nonumber\\
&& \quad (\tilde{\mbox{\boldmath$w$}}_{\mathcal D},\tilde{\mbox{\boldmath$g$}}),\tilde{\mbox{\boldmath$g$}} \in \mathcal{R}_{\mathcal D}, (\tilde{\mbox{\boldmath$w$}}_{\mathcal D},\tilde{\mbox{\boldmath$g$}}) \stackrel{\mathcal S}{=} (\mbox{\boldmath$w$}_{\mathcal D},\mbox{\boldmath$g$}).
\end{eqnarray}

Second, assume that $\tilde{\mbox{\boldmath$w$}}_{\mathcal D}$ is the transmitted message vector for users in $\mathcal D$, and $\tilde{\mbox{\boldmath$g$}}$ is the actual code index vector, with $\tilde{\mbox{\boldmath$g$}} \notin \mathcal{R}_{\mathcal D}$. Define $P_{i[\tilde{\mbox{\scriptsize\boldmath$g$}},\mbox{\scriptsize\boldmath$g$},\mathcal{S}]}$ as the probability that the decoder finds a codeword vector $(\mbox{\boldmath$w$}_{\mathcal D},\mbox{\boldmath$g$})$ with $(\mbox{\boldmath$w$}_{\mathcal D},\mbox{\boldmath$g$}) \stackrel{\mathcal S}{=} (\tilde{\mbox{\boldmath$w$}}_{\mathcal D},\tilde{\mbox{\boldmath$g$}})$ and $\mbox{\boldmath$g$} \in \mathcal{R}_{\mathcal D}$, such that its weighted likelihood is larger than the corresponding typicality threshold,
\begin{eqnarray}
\label{ProofPiCC}
&& P_{i[\tilde{\mbox{\scriptsize\boldmath$g$}},\mbox{\scriptsize\boldmath$g$},\mathcal{S}]} = \nonumber \\
&& Pr \left\{ P(\mbox{\boldmath $y$}|\mbox{\boldmath $x$}_{(\mbox{\scriptsize\boldmath$w$}_{\mathcal D},\mbox{\scriptsize\boldmath$g$}_{\mathcal D})}, \mbox{\boldmath$g$}_{\bar{\mathcal D}}) > e^{-N[\tau_{(\mbox{\tiny\boldmath$g$},\mathcal{S})}(\mbox{\scriptsize\boldmath$x$}_{{\mathcal S} \cap {\mathcal D} }, \mbox{\scriptsize\boldmath$y$})-\alpha(\mbox{\scriptsize\boldmath$g$})]}\right\},\nonumber\\
&& \quad (\mbox{\boldmath$w$}_{\mathcal D},\mbox{\boldmath$g$}),\mbox{\boldmath$g$} \in \mathcal{R}_{\mathcal D}, (\mbox{\boldmath$w$}_{\mathcal D},\mbox{\boldmath$g$}) \stackrel{\mathcal S}{=} (\tilde{\mbox{\boldmath$w$}}_{\mathcal D},\tilde{\mbox{\boldmath$g$}}).
\end{eqnarray}

With the above probability definitions, by applying the union bound over all $\mathcal{S}$, we can upper-bound $\mbox{GEP}_{\mathcal D}(\alpha)$ by
\begin{eqnarray}
\label{ProofPes}
&& \mbox{GEP}_{\mathcal D}(\alpha) \le \frac{1}{\sum_{\mbox{\scriptsize\boldmath$g$}}\exp\left(-N\alpha(\mbox{\boldmath$g$})\right)  }\left\{\sum_{\mbox{\scriptsize\boldmath$g$}\in \mathcal{R}_{\mathcal D}} e^{-N\alpha(\mbox{\scriptsize\boldmath$g$})} \right.     \nonumber \\
&& \quad \sum_{\tiny \begin{array}{c}\mathcal{S}\subset \{1,\cdots,K+M\} \\ \mathcal{D}\setminus\mathcal{S}\ne \emptyset \end{array}} \left[ P_{t[\mbox{\scriptsize\boldmath$g$},\mathcal{S}]} + \sum_{\tilde{\mbox{\scriptsize\boldmath$g$}} \in \mathcal{R}_{\mathcal D},\tilde{\mbox{\scriptsize\boldmath$g$}}_{\mathcal{S}} = \mbox{\scriptsize\boldmath$g$}_{\mathcal{S}}} P_{m[\mbox{\scriptsize\boldmath$g$},\tilde{\mbox{\scriptsize\boldmath$g$}},\mathcal{S}]} \right] \nonumber \\
&& + \sum_{\tilde{\mbox{\scriptsize\boldmath$g$}} \notin \mathcal{R}_{\mathcal D}}e^{-N\alpha(\tilde{\mbox{\scriptsize\boldmath$g$}})} \sum_{\tiny \begin{array}{c}\mathcal{S}\subset \{1,\cdots,K+M\} \\ \mathcal{D}\setminus\mathcal{S}\ne \emptyset \end{array}} \nonumber \\
&& \quad \left. \sum_{\mbox{\scriptsize\boldmath$g$} \in \mathcal{R}_{\mathcal D},\mbox{\scriptsize\boldmath$g$}_{\mathcal{S}} = \tilde{\mbox{\scriptsize\boldmath$g$}}_{\mathcal{S}}} P_{i[\tilde{\mbox{\scriptsize\boldmath$g$}},\mbox{\scriptsize\boldmath$g$},\mathcal{S}]} \right\},
\end{eqnarray}
Next, we will derive individual upper-bounds for each of the terms on the right hand side of (\ref{ProofPes}).

{\bf Step I: } Upper-bounding $P_{m[\mbox{\scriptsize\boldmath$g$},\tilde{\mbox{\scriptsize\boldmath$g$}},\mathcal{S}]}e^{-N\alpha(\mbox{\scriptsize\boldmath$g$})}$

Denote $E_{\mbox{\scriptsize\boldmath $\theta$}}$ as the expectation operator over random variable $\mbox{\boldmath $\theta$}$ which is defined in Section \ref{SectionII}. Given $\mbox{\boldmath$g$},\tilde{\mbox{\boldmath$g$}} \in \mathcal{R}_{\mathcal D}$, $P_{m[\mbox{\scriptsize\boldmath$g$},\tilde{\mbox{\scriptsize\boldmath$g$}},\mathcal{S}]} $ defined in (\ref{ProofPmMC}) can be rewritten as
\begin{eqnarray}
\label{PmBound1}
&& P_{m[\mbox{\scriptsize\boldmath$g$},\tilde{\mbox{\scriptsize\boldmath$g$}},\mathcal{S}]} =  \nonumber \\
&& \quad E_{\mbox{\scriptsize\boldmath $\theta$}_{\mathcal D}} \left[\sum_{\mbox{\scriptsize\boldmath$y$}} P(\mbox{\boldmath $y$}|\mbox{\boldmath $x$}_{(\mbox{\scriptsize\boldmath$w$}_{\mathcal D},\mbox{\scriptsize\boldmath$g$}_{\mathcal D})}, \mbox{\boldmath$g$}_{\bar{\mathcal D}}) \phi_{m[\mbox{\scriptsize\boldmath$g$},\tilde{\mbox{\scriptsize\boldmath$g$}},\mathcal{S}]} (\mbox{\boldmath$x$}_{{\mathcal S} \cap {\mathcal D} }, \mbox{\boldmath$y$}) \right], \nonumber \\
\end{eqnarray}
where $\phi_{m[\mbox{\scriptsize\boldmath$g$},\tilde{\mbox{\scriptsize\boldmath$g$}},\mathcal{S}]} (\mbox{\boldmath$x$}_{{\mathcal S} \cap {\mathcal D} }, \mbox{\boldmath$y$}) = 1$ if $P(\mbox{\boldmath $y$}|\mbox{\boldmath $x$}_{(\mbox{\scriptsize\boldmath$w$}_{\mathcal D},\mbox{\scriptsize\boldmath$g$}_{\mathcal D})}, \mbox{\boldmath$g$}_{\bar{\mathcal D}}) e^{-N\alpha(\mbox{\scriptsize\boldmath$g$})} \le P(\mbox{\boldmath $y$}|\mbox{\boldmath $x$}_{(\tilde{\mbox{\scriptsize\boldmath$w$}}_{\mathcal D}, \tilde{\mbox{\scriptsize\boldmath$g$}}_{\mathcal D})}, \tilde{\mbox{\boldmath$g$}}_{\bar{\mathcal D}})e^{-N\alpha(\tilde{\mbox{\scriptsize\boldmath$g$}})} $ for some $(\tilde{\mbox{\boldmath$w$}}_{\mathcal D},\tilde{\mbox{\boldmath$g$}})$ with $\tilde{\mbox{\boldmath$g$}} \in \mathcal{R}_{\mathcal D}, (\tilde{\mbox{\boldmath$w$}}_{\mathcal D},\tilde{\mbox{\boldmath$g$}}) \stackrel{\mathcal S}{=} (\mbox{\boldmath$w$}_{\mathcal D},\mbox{\boldmath$g$})$. Otherwise, $\phi_{m[\mbox{\scriptsize\boldmath$g$},\tilde{\mbox{\scriptsize\boldmath$g$}},\mathcal{S}]} (\mbox{\boldmath$x$}_{{\mathcal S} \cap {\mathcal D} }, \mbox{\boldmath$y$}) = 0$. We can upper-bound $\phi_{m[\mbox{\scriptsize\boldmath$g$},\tilde{\mbox{\scriptsize\boldmath$g$}},\mathcal{S}]} (\mbox{\boldmath$x$}_{{\mathcal S} \cap {\mathcal D} }, \mbox{\boldmath$y$}) $ for any constants $\rho>0$ and $s>0$ as follows,
\begin{eqnarray}
\label{PhiBoundPmCC}
&& \phi_{m[\mbox{\scriptsize\boldmath$g$},\tilde{\mbox{\scriptsize\boldmath$g$}},\mathcal{S}]} (\mbox{\boldmath$x$}_{{\mathcal S} \cap {\mathcal D} }, \mbox{\boldmath$y$}) \le   e^{-Ns(\alpha(\tilde{\mbox{\scriptsize\boldmath$g$}})-\alpha(\mbox{\scriptsize\boldmath$g$}))} \nonumber \\
&& \left\{ \frac{\sum_{\tilde{\mbox{\scriptsize\boldmath$w$}}_{\mathcal D}, (\tilde{\mbox{\scriptsize\boldmath$w$}}_{\mathcal D},\tilde{\scriptsize\mbox{\boldmath$g$}}) \stackrel{\mathcal S}{=} (\mbox{\scriptsize\boldmath$w$}_{\mathcal D},\mbox{\scriptsize\boldmath$g$})} P(\mbox{\boldmath $y$}|\mbox{\boldmath $x$}_{(\tilde{\mbox{\scriptsize\boldmath$w$}}_{\mathcal D},\tilde{\mbox{\scriptsize\boldmath$g$}}_{\mathcal D})}, \tilde{\mbox{\boldmath$g$}}_{\bar{\mathcal D}} )^{\frac{s}{\rho}} } {P(\mbox{\boldmath $y$}|\mbox{\boldmath $x$}_{(\mbox{\scriptsize\boldmath$w$}_{\mathcal D},\mbox{\scriptsize\boldmath$g$}_{\mathcal D})}, \mbox{\boldmath$g$}_{\bar{\mathcal D}})^{\frac{s}{\rho}}}  \right\}^{\rho}.
\end{eqnarray}
Substituting (\ref{PhiBoundPmCC}) back into (\ref{PmBound1}) gives,
\begin{eqnarray}
\label{PmBound2}
&& P_{m[\mbox{\scriptsize\boldmath$g$},\tilde{\mbox{\scriptsize\boldmath$g$}},\mathcal{S}]} \le \nonumber \\
&& \quad e^{-Ns(\alpha(\tilde{\mbox{\scriptsize\boldmath$g$}})-\alpha(\mbox{\scriptsize\boldmath$g$}))} E_{\mbox{\scriptsize\boldmath $\theta$}_{\mathcal D}} \Biggl[ \sum_{\mbox{\scriptsize\boldmath$y$}} P(\mbox{\boldmath $y$}|\mbox{\boldmath $x$}_{(\mbox{\scriptsize\boldmath$w$}_{\mathcal D},\mbox{\scriptsize\boldmath$g$}_{\mathcal D})},\mbox{\boldmath$g$}_{\bar{\mathcal D}})   \nonumber \\
&&  \times\left.    \left\{ \frac{\sum_{\tilde{\mbox{\scriptsize\boldmath$w$}}_{\mathcal D}, (\tilde{\mbox{\scriptsize\boldmath$w$}}_{\mathcal D},\tilde{\scriptsize\mbox{\boldmath$g$}}) \stackrel{\mathcal S}{=} (\mbox{\scriptsize\boldmath$w$}_{\mathcal D},\mbox{\scriptsize\boldmath$g$})} P(\mbox{\boldmath $y$}|\mbox{\boldmath $x$}_{(\tilde{\mbox{\scriptsize\boldmath$w$}}_{\mathcal D},\tilde{\mbox{\scriptsize\boldmath$g$}}_{\mathcal D})}, \tilde{\mbox{\boldmath$g$}}_{\bar{\mathcal D}} )^{\frac{s}{\rho}} } {P(\mbox{\boldmath $y$}|\mbox{\boldmath $x$}_{(\mbox{\scriptsize\boldmath$w$}_{\mathcal D},\mbox{\scriptsize\boldmath$g$}_{\mathcal D})}, \mbox{\boldmath$g$}_{\bar{\mathcal D}})^{\frac{s}{\rho}} }  \right\}^{\rho}
\right] \nonumber\\
&& =  e^{-Ns(\alpha(\tilde{\mbox{\scriptsize\boldmath$g$}})-\alpha(\mbox{\scriptsize\boldmath$g$}))} \sum_{\mbox{\scriptsize\boldmath$y$}} E_{\mbox{\scriptsize\boldmath $\theta$}_{\mathcal{S}\cap\mathcal{D}}} \Biggl[        \nonumber \\
&& \quad  E_{\mbox{\scriptsize\boldmath $\theta$}_{\mathcal D\setminus \mathcal{S}}} \left[  P(\mbox{\boldmath $y$}|\mbox{\boldmath $x$}_{(\mbox{\scriptsize\boldmath$w$}_{\mathcal D},\mbox{\scriptsize\boldmath$g$}_{\mathcal D})},\mbox{\boldmath$g$}_{\bar{\mathcal D}}) ^{1-s} \right] E_{\mbox{\scriptsize\boldmath $\theta$}_{\mathcal D\setminus \mathcal{S}}} \Biggl [ \nonumber \\
&& \left. \left.\left\{ \sum_{\tilde{\mbox{\scriptsize\boldmath$w$}}_{\mathcal D}, (\tilde{\mbox{\scriptsize\boldmath$w$}}_{\mathcal D},\tilde{\scriptsize\mbox{\boldmath$g$}}) \stackrel{\mathcal S}{=} (\mbox{\scriptsize\boldmath$w$}_{\mathcal D},\mbox{\scriptsize\boldmath$g$})} P(\mbox{\boldmath $y$}|\mbox{\boldmath $x$}_{(\tilde{\mbox{\scriptsize\boldmath$w$}}_{\mathcal D},\tilde{\mbox{\scriptsize\boldmath$g$}}_{\mathcal D})},\tilde{\mbox{\boldmath$g$}}_{\bar{\mathcal D}})^{\frac{s}{\rho}}  \right\}^{\rho} \right]\right].
\end{eqnarray}
The second step in (\ref{PmBound2}) is due to independence between the codewords corresponding to $(\mbox{\boldmath$w$}_{\mathcal D\setminus \mathcal{S}},\mbox{\boldmath$g$}_{\mathcal D\setminus \mathcal{S}})$ and $(\tilde{\mbox{\boldmath$w$}}_{\mathcal D\setminus\mathcal{S}},\tilde{\mbox{\boldmath$g$}}_{\mathcal D\setminus\mathcal{S}})$.

With the assumption of $0<\rho\le 1$, we can further bound $P_{m[\mbox{\scriptsize\boldmath$g$},\tilde{\mbox{\scriptsize\boldmath$g$}},\mathcal{S}]} $ by
\begin{eqnarray}
\label{PmBound3}
&&P_{m[\mbox{\scriptsize\boldmath$g$},\tilde{\mbox{\scriptsize\boldmath$g$}},\mathcal{S}]} \le e^{-Ns(\alpha(\tilde{\mbox{\scriptsize\boldmath$g$}})-\alpha(\mbox{\scriptsize\boldmath$g$}))} e^{N\rho \sum_{k\in\mathcal D\setminus\mathcal{S}} \tilde{r}_k}    \nonumber \\
&& \quad \times \sum_{\mbox{\scriptsize\boldmath$y$}} E_{\mbox{\scriptsize\boldmath $\theta$}_{\mathcal{S}\cap\mathcal{D}}} \Biggl[ E_{\mbox{\scriptsize\boldmath $\theta$}_{\mathcal D \setminus \mathcal{S}}} \left[ P(\mbox{\boldmath $y$}|\mbox{\boldmath $x$}_{(\mbox{\scriptsize\boldmath$w$}_{\mathcal D},\mbox{\scriptsize\boldmath$g$}_{\mathcal D})},\mbox{\boldmath$g$}_{\bar{\mathcal D}}) ^{1-s} \right]     \nonumber \\
&& \quad \times \left\{ E_{\mbox{\scriptsize\boldmath $\theta$}_{\mathcal D\setminus \mathcal{S}}} \left[  P(\mbox{\boldmath $y$}|\mbox{\boldmath $x$}_{(\tilde{\mbox{\scriptsize\boldmath$w$}}_{\mathcal D},\tilde{\mbox{\scriptsize\boldmath$g$}}_{\mathcal D})},\tilde{\mbox{\boldmath$g$}}_{\bar{\mathcal D}})^{\frac{s}{\rho}}  \right]\right\}^{\rho} \Biggr],
\end{eqnarray}
where $\tilde{r}_k$ is the communication rate of code $\tilde{g}_k$.

The bound in (\ref{PmBound3}) holds for all $0<\rho\le 1$ and $s>0$, and becomes trivial for $s>1$. Consequently, (\ref{PmBound3}) gives the following upper bound,
\begin{eqnarray}
\label{PmBound4}
P_{m[\mbox{\scriptsize\boldmath$g$},\tilde{\mbox{\scriptsize\boldmath$g$}},\mathcal{S}]}e^{-N\alpha(\mbox{\scriptsize\boldmath$g$})} \le \exp \left\{ -NE_{m \mathcal{D}}(\mathcal{S}, \mbox{\boldmath$g$}, \tilde{\mbox{\boldmath$g$}}) \right\},
\end{eqnarray}
where $E_{m \mathcal{D}}(\mathcal{S}, \mbox{\boldmath$g$}, \tilde{\mbox{\boldmath$g$}}) $ is specified in (\ref{EmEiDDcoder}).

{\bf Step II: } Upper-bounding $P_{t[\mbox{\scriptsize\boldmath$g$},\mathcal{S}]}e^{-N\alpha(\mbox{\scriptsize\boldmath$g$})}$

Given that $\mbox{\boldmath$g$}\in \mathcal{R}_{\mathcal D}$, we can rewrite $P_{t[\mbox{\scriptsize\boldmath$g$},\mathcal{S}]} $, defined in (\ref{ProofPtCC}), as follows,
\begin{eqnarray}
\label{PtBound1}
&& P_{t[\mbox{\scriptsize\boldmath$g$},\mathcal{S}]} = \nonumber \\
&& \quad E_{\mbox{\scriptsize\boldmath $\theta$}_{\mathcal D}} \left[\sum_{\mbox{\scriptsize\boldmath$y$}} P(\mbox{\boldmath $y$}|\mbox{\boldmath $x$}_{(\mbox{\scriptsize\boldmath$w$}_{\mathcal D},\mbox{\scriptsize\boldmath$g$}_{\mathcal D})},\mbox{\boldmath$g$}_{\bar{\mathcal D}}) \phi_{t[\mbox{\scriptsize\boldmath$g$},\mathcal{S}]} (\mbox{\boldmath$x$}_{{\mathcal S} \cap {\mathcal D} }, \mbox{\boldmath$y$})\right], \nonumber \\
\end{eqnarray}
where $\phi_{t[\mbox{\scriptsize\boldmath$g$},\mathcal{S}]} (\mbox{\boldmath$x$}_{{\mathcal S} \cap {\mathcal D} }, \mbox{\boldmath$y$}) = 1$ if $P(\mbox{\boldmath $y$}|\mbox{\boldmath $x$}_{(\mbox{\scriptsize\boldmath$w$}_{\mathcal D},\mbox{\scriptsize\boldmath$g$}_{\mathcal D})},\mbox{\boldmath$g$}_{\bar{\mathcal D}})e^{-N\alpha(\mbox{\scriptsize\boldmath$g$})} \le e^{-N\tau_{(\mbox{\tiny\boldmath$g$},\mathcal{S})}(\mbox{\scriptsize\boldmath$x$}_{{\mathcal S} \cap {\mathcal D} }, \mbox{\scriptsize\boldmath$y$})}$, otherwise $\phi_{t[\mbox{\scriptsize\boldmath$g$},\mathcal{S}]} (\mbox{\boldmath$x$}_{{\mathcal S} \cap {\mathcal D} }, \mbox{\boldmath$y$}) = 0$. Note that the value of $\tau_{(\mbox{\tiny\boldmath$g$}, \mathcal{S})}(\mbox{\boldmath$x$}_{{\mathcal S} \cap {\mathcal D} }, \mbox{\boldmath$y$})$ will be determined in Step IV. Similarly, we can bound $\phi_{t[\mbox{\scriptsize\boldmath$g$},\mathcal{S}]} (\mbox{\boldmath$x$}_{{\mathcal S} \cap {\mathcal D} }, \mbox{\boldmath$y$})$, for any $s_1>0$, as follows,
\begin{eqnarray}
\phi_{t[\mbox{\scriptsize\boldmath$g$},\mathcal{S}]} (\mbox{\boldmath$x$}_{{\mathcal S} \cap {\mathcal D} }, \mbox{\boldmath$y$}) \le
\frac{e^{-Ns_1\tau_{(\mbox{\tiny\boldmath$g$}, \mathcal{S})}(\mbox{\scriptsize\boldmath$x$}_{{\mathcal S} \cap {\mathcal D} }, \mbox{\scriptsize\boldmath$y$})}}
{P(\mbox{\boldmath $y$}|\mbox{\boldmath $x$}_{(\mbox{\scriptsize\boldmath$w$}_{\mathcal D},\mbox{\scriptsize\boldmath$g$}_{\mathcal D})},\mbox{\boldmath$g$}_{\bar{\mathcal D}})^{s_1}e^{-Ns_1\alpha(\mbox{\scriptsize\boldmath$g$})}}.
\end{eqnarray}
This yields,
\begin{eqnarray}
\label{MCBound4P_t}
&& P_{t[\mbox{\scriptsize\boldmath$g$},\mathcal{S}]}e^{-N\alpha(\mbox{\scriptsize\boldmath$g$})} \le E_{\mbox{\scriptsize\boldmath $\theta$}_{\mathcal D}} \Biggl[   \sum_{\mbox{\scriptsize\boldmath$y$}} P(\mbox{\boldmath $y$}|\mbox{\boldmath $x$}_{(\mbox{\scriptsize\boldmath$w$}_{\mathcal D},\mbox{\scriptsize\boldmath$g$}_{\mathcal D})},\mbox{\boldmath$g$}_{\bar{\mathcal D}})^{1-s_1} \nonumber \\
&& \quad \times \left. e^{-N(1-s_1)\alpha(\mbox{\scriptsize\boldmath$g$})}e^{-Ns_1\tau_{(\mbox{\tiny\boldmath$g$},\mathcal{S})}(\mbox{\scriptsize\boldmath$x$}_{{\mathcal S} \cap {\mathcal D} }, \mbox{\scriptsize\boldmath$y$})} \right] \nonumber\\
&& =  \sum_{\mbox{\scriptsize\boldmath$y$}} E_{\mbox{\scriptsize\boldmath$\theta$}_{\mathcal{S}\cap\mathcal{D}}} \Biggl[  E_{\mbox{\scriptsize\boldmath$\theta$}_{\mathcal D\setminus \mathcal{S}}} \left[ P(\mbox{\boldmath $y$}|\mbox{\boldmath $x$}_{(\mbox{\scriptsize\boldmath$w$}_{\mathcal D},\mbox{\scriptsize\boldmath$g$}_{\mathcal D})},\mbox{\boldmath$g$}_{\bar{\mathcal D}})^{1-s_1} \right]      \nonumber \\
&& \quad \times e^{-N(1-s_1)\alpha(\mbox{\scriptsize\boldmath$g$})}e^{-Ns_1 \tau_{(\mbox{\tiny\boldmath$g$},\mathcal{S})}(\mbox{\scriptsize\boldmath$x$}_{{\mathcal S} \cap {\mathcal D} }, \mbox{\scriptsize\boldmath$y$}) } \Biggr] .
\end{eqnarray}

{\bf Step III:}  Upper-Bounding $P_{i[\tilde{\mbox{\scriptsize\boldmath$g$}},\mbox{\scriptsize\boldmath$g$},\mathcal{S}]}e^{-N\alpha(\tilde{\mbox{\scriptsize\boldmath$g$}})}$

Given $\tilde{\mbox{\boldmath$g$}} \notin \mathcal{R}_{\mathcal D}$ and $\mbox{\boldmath$g$} \in \mathcal{R}_{\mathcal D}$, we rewrite $P_{i[\tilde{\mbox{\scriptsize\boldmath$g$}},\mbox{\scriptsize\boldmath$g$},\mathcal{S}]}$
as
\begin{equation}
\label{P_iBound1}
P_{i[\tilde{\mbox{\scriptsize\boldmath$g$}},\mbox{\scriptsize\boldmath$g$},\mathcal{S}]}
= E_{{\mbox{\scriptsize\boldmath $\theta$}}_{\mathcal D}} \left[\sum_{\mbox{\scriptsize\boldmath$y$}} P(\mbox{\boldmath $y$}|\mbox{\boldmath $x$}_{(\tilde{\mbox{\scriptsize\boldmath$w$}}_{\mathcal D},\tilde{\mbox{\scriptsize\boldmath$g$}}_{\mathcal D})},\tilde{\mbox{\boldmath$g$}}_{\bar{\mathcal D}})) \phi_{[\tilde{\mbox{\scriptsize\boldmath$g$}},\mbox{\scriptsize\boldmath$g$},\mathcal{S}]} (\mbox{\boldmath$x$}_{{\mathcal S} \cap {\mathcal D} }, \mbox{\boldmath$y$}) \right],
\end{equation}
where $\phi_{[\tilde{\mbox{\scriptsize\boldmath$g$}},\mbox{\scriptsize\boldmath$g$},\mathcal{S}]} (\mbox{\boldmath$x$}_{{\mathcal S} \cap {\mathcal D} }, \mbox{\boldmath$y$}) = 1$ if there exists $(\mbox{\boldmath$w$}_{\mathcal D},\mbox{\boldmath$g$})$ with $\mbox{\boldmath$g$}\in \mathcal{R}_{\mathcal D}$ and $(\mbox{\boldmath$w$}_{\mathcal D},\mbox{\boldmath$g$}) \stackrel{\mathcal S}{=} (\tilde{\mbox{\boldmath$w$}}_{\mathcal D},\tilde{\mbox{\boldmath$g$}})$, such that $P(\mbox{\boldmath $y$}|\mbox{\boldmath $x$}_{(\mbox{\scriptsize\boldmath$w$}_{\mathcal D},\mbox{\scriptsize\boldmath$g$}_{\mathcal D})},\mbox{\boldmath$g$}_{\bar{\mathcal D}})e^{-N\alpha(\mbox{\scriptsize\boldmath$g$})} > e^{-N\tau_{(\mbox{\tiny\boldmath$g$},\mathcal{S})}(\mbox{\scriptsize\boldmath$x$}_{{\mathcal S} \cap {\mathcal D} }, \mbox{\scriptsize\boldmath$y$})}$ is satisfied. Otherwise, $\phi_{[\tilde{\mbox{\scriptsize\boldmath$g$}},\mbox{\scriptsize\boldmath$g$},\mathcal{S}]} (\mbox{\boldmath$x$}_{{\mathcal S} \cap {\mathcal D} }, \mbox{\boldmath$y$}) = 0$.

For any $s_2 >0$ and $\tilde{\rho} > 0$, $\phi_{[\tilde{\mbox{\scriptsize\boldmath$g$}},\mbox{\scriptsize\boldmath$g$},\mathcal{S}]} (\mbox{\boldmath$x$}_{{\mathcal S} \cap {\mathcal D} }, \mbox{\boldmath$y$})$
 can be bounded by,
\begin{eqnarray}
\label{PhiBoundPiCC}
&& \phi_{[\tilde{\mbox{\scriptsize\boldmath$g$}},\mbox{\scriptsize\boldmath$g$},\mathcal{S}]} (\mbox{\boldmath$x$}_{{\mathcal S} \cap {\mathcal D} }, \mbox{\boldmath$y$})          \le e^{-Ns_2\alpha(\mbox{\scriptsize\boldmath$g$})} \nonumber \\
&& \times
\left\{ \frac{\sum_{\mbox{\scriptsize\boldmath$w$}_{\mathcal D},(\mbox{\scriptsize\boldmath$w$}_{\mathcal D},\mbox{\scriptsize\boldmath$g$}) \stackrel{\mathcal S}{=} (\tilde{\mbox{\scriptsize\boldmath$w$}}_{\mathcal D},\tilde{\mbox{\scriptsize\boldmath$g$}})} P(\mbox{\boldmath $y$}|\mbox{\boldmath $x$}_{(\mbox{\scriptsize\boldmath$w$}_{\mathcal D},\mbox{\scriptsize\boldmath$g$}_{\mathcal D})},\mbox{\boldmath$g$}_{\bar{\mathcal D}})^{\frac{s_2}{\tilde{\rho}}}}
{e^{-N \frac{s_2} {\tilde{\rho}} \tau_{(\mbox{\tiny\boldmath$g$},\mathcal{S})}(\mbox{\scriptsize\boldmath$x$}_{{\mathcal S} \cap {\mathcal D} }, \mbox{\scriptsize\boldmath$y$})}}\right\} ^{\tilde{\rho}}.
\end{eqnarray}
Substituting (\ref{PhiBoundPiCC}) into (\ref{P_iBound1}) yields,
\begin{eqnarray}
\label{P_iBound2}
&& P_{i[\tilde{\mbox{\scriptsize\boldmath$g$}},\mbox{\scriptsize\boldmath$g$},\mathcal{S}]} \le e^{-Ns_2\alpha(\mbox{\scriptsize\boldmath$g$})} \nonumber\\
&& \quad \times \sum_{\mbox{\scriptsize\boldmath$y$}}  E_{{\mbox{\scriptsize\boldmath $\theta$}}_{\mathcal D}} \Biggl[P(\mbox{\boldmath $y$}|\mbox{\boldmath $x$}_{(\tilde{\mbox{\scriptsize\boldmath$w$}}_{\mathcal D},\tilde{\mbox{\scriptsize\boldmath$g$}}_{\mathcal D})},\tilde{\mbox{\boldmath$g$}}_{\bar{\mathcal D}})e^{N s_2\tau_{(\mbox{\tiny\boldmath$g$},\mathcal{S})}(\mbox{\scriptsize\boldmath$x$}_{{\mathcal S} \cap {\mathcal D} }, \mbox{\scriptsize\boldmath$y$})}       \nonumber \\
&& \quad \times \left.     \left\{ \sum_{\mbox{\scriptsize\boldmath$w$}_{\mathcal D},(\mbox{\scriptsize\boldmath$w$}_{\mathcal D},\mbox{\scriptsize\boldmath$g$}) \stackrel{\mathcal S}{=} (\tilde{\mbox{\scriptsize\boldmath$w$}}_{\mathcal D},\tilde{\mbox{\scriptsize\boldmath$g$}})} P(\mbox{\boldmath $y$}|\mbox{\boldmath $x$}_{(\mbox{\scriptsize\boldmath$w$}_{\mathcal D},\mbox{\scriptsize\boldmath$g$}_{\mathcal D})},\mbox{\boldmath$g$}_{\bar{\mathcal D}})^{\frac{s_2}{\tilde{\rho}}}
\right\} ^{\tilde{\rho}} \right]. \nonumber \\
\end{eqnarray}
The independence between $(\mbox{\boldmath$w$}_{\mathcal D\setminus \mathcal{S}},\mbox{\boldmath$g$}_{\mathcal D\setminus \mathcal{S}})$ and $(\tilde{\mbox{\boldmath$w$}}_{\mathcal D\setminus \mathcal{S}},\tilde{\mbox{\boldmath$g$}}_{\mathcal D\setminus \mathcal{S}})$ allows us to rewrite the above bound as
\begin{eqnarray}
\label{P_iBound3}
&& P_{i[\tilde{\mbox{\scriptsize\boldmath$g$}},\mbox{\scriptsize\boldmath$g$},\mathcal{S}]}
\le \sum_{\mbox{\scriptsize\boldmath$y$}}  E_{{\mbox{\scriptsize\boldmath $\theta$}}_{\mathcal{S}\cap \mathcal D}} \Biggl[ E_{{\mbox{\scriptsize\boldmath $\theta$}}_{\mathcal D \setminus \mathcal{S}}} \left[ P(\mbox{\boldmath $y$}|\mbox{\boldmath $x$}_{(\tilde{\mbox{\scriptsize\boldmath$w$}}_{\mathcal D},\tilde{\mbox{\scriptsize\boldmath$g$}}_{\mathcal D})},\tilde{\mbox{\boldmath$g$}}_{\bar{\mathcal D}}) \right] \nonumber \\
&& \quad \times e^{N s_2(\tau_{(\mbox{\tiny\boldmath$g$},\mathcal{S})}(\mbox{\scriptsize\boldmath$x$}_{{\mathcal S} \cap {\mathcal D} }, \mbox{\scriptsize\boldmath$y$})-\alpha(\mbox{\scriptsize\boldmath$g$}))} E_{{\mbox{\scriptsize\boldmath $\theta$}}_{\mathcal D\setminus \mathcal{S}}} \Biggl[  \Biggl\{       \nonumber \\
&& \quad \left. \left.\left.  \sum_{\mbox{\scriptsize\boldmath$w$}_{\mathcal D},(\mbox{\scriptsize\boldmath$w$}_{\mathcal D},\mbox{\scriptsize\boldmath$g$}) \stackrel{\mathcal S}{=} (\tilde{\mbox{\scriptsize\boldmath$w$}}_{\mathcal D},\tilde{\mbox{\scriptsize\boldmath$g$}})} P(\mbox{\boldmath $y$}|\mbox{\boldmath $x$}_{(\mbox{\scriptsize\boldmath$w$}_{\mathcal D},\mbox{\scriptsize\boldmath$g$}_{\mathcal D})},\mbox{\boldmath$g$}_{\bar{\mathcal D}})^{\frac{s_2}{\tilde{\rho}}}
\right\} ^{\tilde{\rho}}\right] \right].
\end{eqnarray}

With the assumption of $0 < \tilde{\rho} \le 1$, (\ref{P_iBound3}) further implies
\begin{eqnarray}
\label{MCBound4P_i1}
&& P_{i[\tilde{\mbox{\scriptsize\boldmath$g$}},\mbox{\scriptsize\boldmath$g$},\mathcal{S}]} \le \sum_{\mbox{\scriptsize\boldmath$y$}} E_{\mbox{\scriptsize\boldmath$\theta$}_{\mathcal{S}\cap\mathcal D}} \Biggl[  E_{\mbox{\scriptsize\boldmath$\theta$}_{\mathcal D \setminus \mathcal{S}}} \left[ P(\mbox{\boldmath$y$}|\mbox{\boldmath$x$}_{(\tilde{\mbox{\scriptsize\boldmath$w$}}_{\mathcal D},\tilde{\mbox{\scriptsize\boldmath$g$}}_{\mathcal D})},\tilde{\mbox{\boldmath$g$}}_{\bar{\mathcal D}}) \right] \nonumber\\
&& \quad \times \left\{E_{\mbox{\scriptsize\boldmath$\theta$}_{\mathcal D \setminus \mathcal{S}}} \left[ P(\mbox{\boldmath$y$}|\mbox{\boldmath$x$}_{(\mbox{\scriptsize\boldmath$w$}_{\mathcal D},\mbox{\scriptsize\boldmath$g$}_{\mathcal D})},\mbox{\boldmath$g$}_{\bar{\mathcal D}})^{\frac{s_2}{\tilde{\rho}}} \right] \right\}^{\tilde{\rho}}           \nonumber \\
&& \quad \times          \left. e^{Ns_2 (\tau_{(\mbox{\tiny\boldmath$g$},\mathcal{S})}(\mbox{\scriptsize\boldmath$x$}_{{\mathcal S} \cap {\mathcal D} }, \mbox{\scriptsize\boldmath$y$})-\alpha(\mbox{\scriptsize\boldmath$g$}))}e^{N\tilde{\rho}\sum_{k\in\mathcal D \setminus\mathcal{S}}r_k} \right].
\end{eqnarray}
Consequently, we can upper-bound $P_{i[\tilde{\mbox{\scriptsize\boldmath$g$}},\mbox{\scriptsize\boldmath$g$},\mathcal{S}]}e^{-N\alpha(\tilde{\mbox{\scriptsize\boldmath$g$}})}$ by
\begin{eqnarray}
\label{MCBound4P_i}
&& P_{i[\tilde{\mbox{\scriptsize\boldmath$g$}},\mbox{\scriptsize\boldmath$g$},\mathcal{S}]}e^{-N\alpha(\tilde{\mbox{\scriptsize\boldmath$g$}})} \le \max_{\mbox{\scriptsize\boldmath$g$}' \notin \mathcal{R}_{\mathcal D},\mbox{\scriptsize\boldmath$g$}'_{\mathcal{S}} = \mbox{\scriptsize\boldmath$g$}_{\mathcal{S}}} e^{-N\alpha(\mbox{\scriptsize\boldmath$g$}')} \nonumber \\
&& \times \sum_{\mbox{\scriptsize\boldmath$y$}}   E_{\mbox{\scriptsize\boldmath$\theta$}_{\mathcal{S}\cap \mathcal D}} \Bigl[  E_{\mbox{\scriptsize\boldmath$\theta$}_{\mathcal D \setminus \mathcal{S}}} \left[ P(\mbox{\boldmath$y$}|\mbox{\boldmath$x$}_{(\mbox{\scriptsize\boldmath$w$}'_{\mathcal D},\mbox{\scriptsize\boldmath$g$}'_{\mathcal D})},\mbox{\boldmath$g$}'_{\bar{\mathcal D}}) \right]  \nonumber\\
&& \quad \times \left\{ E_{\mbox{\scriptsize\boldmath$\theta$}_{\mathcal D \setminus \mathcal{S}}} \left[ P(\mbox{\boldmath$y$}|\mbox{\boldmath$x$}_{(\mbox{\scriptsize\boldmath$w$}_{\mathcal D},\mbox{\scriptsize\boldmath$g$}_{\mathcal D})},\mbox{\boldmath$g$}_{\bar{\mathcal D}})^{\frac{s_2}{\tilde{\rho}}} \right] \right\}^{\tilde{\rho}}      \nonumber \\
&& \quad \times e^{Ns_2 (\tau_{(\mbox{\tiny\boldmath$g$},\mathcal{S})}(\mbox{\scriptsize\boldmath$x$}_{{\mathcal S} \cap {\mathcal D} }, \mbox{\scriptsize\boldmath$y$})-\alpha(\mbox{\scriptsize\boldmath$g$})) }   \left. e^{N\tilde{\rho}\sum_{k\in\mathcal D\setminus\mathcal{S}}r_k} \right].
\end{eqnarray}
Note that the upper bound in (\ref{MCBound4P_i}) is not a function of $\tilde{\mbox{\boldmath$g$}}_{\bar{\mathcal{S}}}$.

{\bf Step IV: } Choosing $\tau_{(\mbox{\scriptsize\boldmath$g$},\mathcal{S})}(\mbox{\boldmath$x$}_{{\mathcal S} \cap {\mathcal D} }, \mbox{\boldmath$y$})$

Let $\tilde{\mbox{\boldmath$g$}}^*\not\in\mathcal R_{\mathcal D}$ be the code index vector that maximizes the right hand side of (\ref{MCBound4P_i}). Given $\mbox{\boldmath$g$}\in \mathcal{R}_{\mathcal D}$, $\mbox{\boldmath$y$}$ and auxiliary variables $s_1>0$, $s_2>0$, $0< \tilde{\rho} \le 1$, we choose  $\tau_{(\mbox{\scriptsize\boldmath$g$},\mathcal{S})}(\mbox{\boldmath$x$}_{{\mathcal S} \cap {\mathcal D} }, \mbox{\boldmath$y$})$ such that the following equality is satisfied,
\begin{eqnarray}
\label{TauEquality}
&& E_{\mbox{\scriptsize\boldmath$\theta$}_{\mathcal D\setminus \mathcal{S}}} \left[ P(\mbox{\boldmath $y$}|\mbox{\boldmath $x$}_{(\mbox{\scriptsize\boldmath$w$}_{\mathcal D},\mbox{\scriptsize\boldmath$g$}_{\mathcal D})},\mbox{\boldmath$g$}_{\bar{\mathcal D}})^{1-s_1} e^{-N(1-s_1)\alpha(\mbox{\scriptsize\boldmath$g$})}\right]  \nonumber \\
&& \times  e^{-Ns_1 \tau_{(\mbox{\tiny\boldmath$g$},\mathcal{S})}(\mbox{\scriptsize\boldmath$x$}_{{\mathcal S} \cap {\mathcal D} }, \mbox{\scriptsize\boldmath$y$}) } =      \nonumber \\
&& E_{\mbox{\scriptsize\boldmath$\theta$}_{\mathcal D \setminus \mathcal{S}}} \left[ P(\mbox{\boldmath$y$}|\mbox{\boldmath$x$}_{(\tilde{\mbox{\scriptsize\boldmath$w$}}^*_{\mathcal D},\tilde{\mbox{\scriptsize\boldmath$g$}}^*_{\mathcal D})},\tilde{\mbox{\boldmath$g$}}^*_{\bar{\mathcal D}})e^{-N\alpha(\mbox{\scriptsize\boldmath$g$}^*)} \right] \nonumber \\
&& \quad \times  \left\{ E_{\mbox{\scriptsize\boldmath$\theta$}_{\mathcal D \setminus \mathcal{S}}} \left[ P(\mbox{\boldmath$y$}|\mbox{\boldmath$x$}_{(\mbox{\scriptsize\boldmath$w$}_{\mathcal D},\mbox{\scriptsize\boldmath$g$}_{\mathcal D})},\mbox{\boldmath$g$}_{\bar{\mathcal D}})^{\frac{s_2}{\tilde{\rho}}}e^{-N\frac{s_2}{\tilde{\rho}}\alpha(\mbox{\scriptsize\boldmath$g$})} \right] \right\}^{\tilde{\rho}}         \nonumber \\
&& \quad \times e^{Ns_2 \tau_{(\mbox{\tiny\boldmath$g$},\mathcal{S})}(\mbox{\scriptsize\boldmath$x$}_{{\mathcal S} \cap {\mathcal D} }, \mbox{\scriptsize\boldmath$y$})}   e^{N\tilde{\rho}\sum_{k\in\mathcal D\setminus\mathcal{S}}r_k} .
\end{eqnarray}
Note that here we do not enforce the natural constraint that $\tau_{(\mbox{\scriptsize\boldmath$g$},\mathcal{S})}(\mbox{\boldmath$x$}_{{\mathcal S} \cap {\mathcal D} }, \mbox{\boldmath$y$})\ge 0$, and hence finding a solution for (\ref{TauEquality}) is always possible. The solution gives the desired typicality threshold, denoted by $\tau_{(\mbox{\scriptsize\boldmath$g$},\mathcal{S})}^*(\mbox{\boldmath$x$}_{{\mathcal S} \cap {\mathcal D} }, \mbox{\boldmath$y$})$, which satisfies
\begin{eqnarray}
\label{TauExpression}
&& e^{-N\tau_{(\mbox{\tiny\boldmath$g$},\mathcal{S})}^*(\mbox{\scriptsize\boldmath$x$}_{{\mathcal S} \cap {\mathcal D} }, \mbox{\scriptsize\boldmath$y$})} = \nonumber \\
&& {\left\{  E_{\mbox{\scriptsize\boldmath$\theta$}_{\mathcal D \setminus\mathcal{S}}} \left[ \left[P(\mbox{\boldmath $y$}|\mbox{\boldmath $x$}_{(\mbox{\scriptsize\boldmath$w$}_{\mathcal D},\mbox{\scriptsize\boldmath$g$}_{\mathcal D})},\mbox{\boldmath$g$}_{\bar{\mathcal D}})e^{-N\alpha(\mbox{\scriptsize\boldmath$g$})} \right]^{1-s_1} \right] \right\} ^{-\frac{1}{s_1+s_2}}}      \nonumber \\
&& \times \left\{ E_{\mbox{\scriptsize\boldmath$\theta$}_{\mathcal D\setminus\mathcal{S}}} \left[ \left[P(\mbox{\boldmath$y$}|\mbox{\boldmath$x$}_{(\mbox{\scriptsize\boldmath$w$}_{\mathcal D},\mbox{\scriptsize\boldmath$g$}_{\mathcal D})},\mbox{\boldmath$g$}_{\bar{\mathcal D}})e^{-N\alpha(\mbox{\scriptsize\boldmath$g$})} \right]^{\frac{s_2}{\tilde{\rho}}} \right] \right\}^{\frac{\tilde{\rho}}{s_1+s_2}}      \nonumber \\
&& \times  \left\{E_{\mbox{\scriptsize\boldmath$\theta$}_{\mathcal D\setminus \mathcal{S}}} \left[ P(\mbox{\boldmath$y$}|\mbox{\boldmath$x$}_{(\tilde{\mbox{\scriptsize\boldmath$w$}}^*_{\mathcal D},\tilde{\mbox{\scriptsize\boldmath$g$}}^*_{\mathcal D})},\tilde{\mbox{\boldmath$g$}}^*_{\bar{\mathcal D}})e^{-N\alpha(\mbox{\scriptsize\boldmath$g$}^*)} \right]\right\}^{\frac{1}{s_1+s_2}} \nonumber\\
&& \times e^{N\frac{\tilde{\rho}}{s_1+s_2}\sum_{k\in\mathcal D\setminus\mathcal{S}}r_k}.
\end{eqnarray}

Substituting (\ref{TauExpression}) into (\ref{MCBound4P_t}), we get
\begin{eqnarray}
\label{MCBound4P_tA}
&& P_{t[\mbox{\scriptsize\boldmath$g$},\mathcal{S}]}e^{-N\alpha(\mbox{\scriptsize\boldmath$g$})}  \le  \sum_{\mbox{\scriptsize\boldmath$y$}} E_{\mbox{\scriptsize\boldmath$\theta$}_{\mathcal{S}\cap \mathcal D}} \Biggl[  \nonumber \\
&& \left\{ E_{\mbox{\scriptsize\boldmath$\theta$}_{\mathcal D\setminus \mathcal{S}}} \left[ \left[P(\mbox{\boldmath $y$}|\mbox{\boldmath $x$}_{(\mbox{\scriptsize\boldmath$w$}_{\mathcal D},\mbox{\scriptsize\boldmath$g$}_{\mathcal D})},\mbox{\boldmath$g$}_{\bar{\mathcal D}})e^{-N\alpha(\mbox{\scriptsize\boldmath$g$})} \right]^{1-s_1} \right]\right\}^{\frac{s_2}{s_1+s_2}}       \nonumber \\
&& \times \left\{ E_{\mbox{\scriptsize\boldmath$\theta$}_{\mathcal D\setminus \mathcal{S}}} \left[ \left[P(\mbox{\boldmath$y$}|\mbox{\boldmath$x$}_{(\mbox{\scriptsize\boldmath$w$}_{\mathcal D},\mbox{\scriptsize\boldmath$g$}_{\mathcal D})},\mbox{\boldmath$g$}_{\bar{\mathcal D}})e^{-N\alpha(\mbox{\scriptsize\boldmath$g$})} \right]^{\frac{s_2}{\tilde{\rho}}} \right] \right\}^{\frac{s_1\tilde{\rho}}{s_1+s_2}}    \nonumber \\
&& \times \left\{E_{\mbox{\scriptsize\boldmath$\theta$}_{\mathcal D \setminus \mathcal{S}}} \left[ P(\mbox{\boldmath$y$}|\mbox{\boldmath$x$}_{(\tilde{\mbox{\scriptsize\boldmath$w$}}^*_{\mathcal D},\tilde{\mbox{\scriptsize\boldmath$g$}}^*_{\mathcal D})},\tilde{\mbox{\boldmath$g$}}^*_{\bar{\mathcal D}})e^{-N\alpha(\mbox{\scriptsize\boldmath$g$}^*)} \right]\right\}^{\frac{s_1}{s_1+s_2}} \nonumber\\
&& \times \left. e^{N\frac{s_1\tilde{\rho}}{s_1+s_2}\sum_{k\in\mathcal D\setminus\mathcal{S}}r_k} \right] .
\end{eqnarray}
Let $s_2 < \tilde{\rho}$ and $s_1 = 1 - \frac{s_2}{\tilde{\rho}}$, and then do a variable change with $\rho = \frac{\tilde{\rho}(\tilde{\rho}-s_2)}{\tilde{\rho} - (1-\tilde{\rho})s_2}$ and $s = 1- \frac{\tilde{\rho}-s_2}{\tilde{\rho} - (1-\tilde{\rho})s_2}$. Inequality (\ref{MCBound4P_tA}) becomes,
\begin{eqnarray}
\label{MCBound4P_tB}
&&P_{t[\mbox{\scriptsize\boldmath$g$},\mathcal{S}]}e^{-N\alpha(\mbox{\scriptsize\boldmath$g$})}   \le  e^{N\rho \sum_{k\in \mathcal D \setminus\mathcal{S}}r_k} \Biggl\{        \nonumber \\
&&  \times \sum_Y \sum_{\mbox{\scriptsize \boldmath $X$}_{\mathcal{S}\cap \mathcal D}}\prod_{k\in \mathcal{S}\cap \mathcal D} P_{X|g_k}(X_k) \left(\sum_{\mbox{\scriptsize \boldmath $X$}_{\mathcal D\setminus \mathcal{S}}}\prod_{k \in \mathcal D\setminus \mathcal{S}}P_{X|g_k}(X_k)\right. \nonumber \\
&&  \times \left.\left[P(Y|\mbox{\boldmath $X$}_{\mathcal D}, \mbox{\boldmath $g$}_{\bar{\mathcal D}})e^{-\alpha(\mbox{\scriptsize\boldmath$g$})}\right]^{\frac{s}{s+\rho}} \right)^{s+\rho} \left(\sum_{\mbox{\scriptsize \boldmath $X$}_{\mathcal D\setminus \mathcal{S}}}\prod_{k \in \mathcal D \setminus \mathcal{S}}\right.     \nonumber \\
&&  \left. \left.P_{X|\tilde{g}^*_k}(X_k)P(Y|\mbox{\boldmath $X$}, \tilde{\mbox{\boldmath $g$}}^*_{\bar{\mathcal D}})e^{-\alpha(\mbox{\scriptsize\boldmath$g$}^*)}  \right)^{1-s}\right\}^{N}.
\end{eqnarray}
Similarly, we get from (\ref{MCBound4P_i}) that
\begin{eqnarray}
\label{MCBound4P_iB}
&&P_{i[\tilde{\mbox{\scriptsize\boldmath$g$}},\mbox{\scriptsize\boldmath$g$},\mathcal{S}]}e^{-N\alpha(\tilde{\mbox{\scriptsize\boldmath$g$}})}   \le   e^{N\rho \sum_{k\in \mathcal D \setminus\mathcal{S}}r_k} \Biggl\{        \nonumber \\
&&  \times \sum_Y \sum_{\mbox{\scriptsize \boldmath $X$}_{\mathcal{S}\cap \mathcal D}}\prod_{k\in \mathcal{S}\cap \mathcal D} P_{X|g_k}(X_k) \left(\sum_{\mbox{\scriptsize \boldmath $X$}_{\mathcal D\setminus \mathcal{S}}}\prod_{k \in \mathcal D\setminus \mathcal{S}}P_{X|g_k}(X_k)\right. \nonumber \\
&&  \times \left.\left[P(Y|\mbox{\boldmath $X$}_{\mathcal D}, \mbox{\boldmath $g$}_{\bar{\mathcal D}})e^{-\alpha(\mbox{\scriptsize\boldmath$g$})}\right]^{\frac{s}{s+\rho}} \right)^{s+\rho} \left(\sum_{\mbox{\scriptsize \boldmath $X$}_{\mathcal D\setminus \mathcal{S}}}\prod_{k \in \mathcal D \setminus \mathcal{S}}\right.     \nonumber \\
&&  \left. \left.P_{X|\tilde{g}^*_k}(X_k)P(Y|\mbox{\boldmath $X$}, \tilde{\mbox{\boldmath $g$}}^*_{\bar{\mathcal D}})e^{-\alpha(\mbox{\scriptsize\boldmath$g$}^*)}  \right)^{1-s}\right\}^{N}.
\end{eqnarray}

(\ref{MCBound4P_tB}) and (\ref{MCBound4P_iB}) imply that
\begin{eqnarray}
\label{BoundPiPt}
&& P_{t[\mbox{\scriptsize\boldmath$g$},\mathcal{S}]}e^{-N\alpha(\mbox{\scriptsize\boldmath$g$})}\le \max_{\mbox{\scriptsize\boldmath$g$}' \notin \mathcal{R}_{\mathcal D},\mbox{\scriptsize\boldmath$g$}'_{\mathcal{S}} = \mbox{\scriptsize\boldmath$g$}_{\mathcal{S}}} \exp \left\{ -N E_{i \mathcal{D}}(\mathcal{S}, \mbox{\boldmath$g$},\mbox{\boldmath $g$}') \right\}, \nonumber \\
&& P_{i[\tilde{\mbox{\scriptsize\boldmath$g$}},\mbox{\scriptsize\boldmath$g$},\mathcal{S}]}e^{-N\alpha(\tilde{\mbox{\scriptsize\boldmath$g$}})} \le \max_{\tiny \begin{array}{c} \mbox{\scriptsize\boldmath$g$}' \notin \mathcal{R}_{\mathcal D} \\ \mbox{\scriptsize\boldmath$g$}'_{\mathcal{S}} = \mbox{\scriptsize\boldmath$g$}_{\mathcal{S}}\end{array}} \exp \left\{ -N E_{i \mathcal{D}}(\mathcal{S}, \mbox{\boldmath$g$},\mbox{\boldmath $g$}') \right\}, \nonumber \\
\end{eqnarray}
where $E_{i \mathcal{D}}(\mathcal{S}, \mbox{\boldmath$g$},\mbox{\boldmath $g$}') $ is given in (\ref{EmEiDDcoder}).

By substituting (\ref{PmBound4}) and (\ref{BoundPiPt}) into (\ref{ProofPes}), we get the desired result.
\end{proof}

\subsection{Proof of Theorem \ref{Theorem7}}
\label{ProofTheorem7}
\begin{proof}
Because the proof essentially follows the same idea of the proof of Theorem \ref{Theorem3}, we will only present the parts that are different from Appendix \ref{ProofTheorem3}.

Given a user subset $\mathcal S \subset \{1, \cdots, K+M\}$, we define the notation $(\mbox{\boldmath$w$}_{\mathcal D},\mbox{\boldmath$g$}) \stackrel{\mathcal S}{=} (\tilde{\mbox{\boldmath$w$}}_{\mathcal D},\tilde{\mbox{\boldmath$g$}})$ as in (\ref{WsGsEquality}). We assume that the following decoding algorithm is used at the receiver. Given the received channel output symbols $\mbox{\boldmath$y$}$, the receiver estimates the code index vector $\mbox{\boldmath$g$}$. The receiver outputs the messages and code index estimates for users in $\mathcal D$, denoted jointly by $(\mbox{\boldmath$w$}_{\mathcal D}, \mbox{\boldmath$g$}_{\mathcal D})$, if the following three conditions are satisfied simultaneously. First, $\mbox{\boldmath$g$} \in \mathcal R_{\mathcal D}$. Second, (\ref{CompoundCriterionAppendix}) holds for all user subsets $\mathcal{S} \subset \{1,\cdots,K+M\}$ with $\mathcal D \setminus \mathcal S \ne \emptyset$. Third, $ (\mbox{\boldmath$w$}_{\mathcal D}, \mbox{\boldmath$g$}) \in \mathcal{R}_{(\mathcal{S},\mbox{\scriptsize \boldmath $y$})}$ for all user subsets $\mathcal{S} \subset \{1,\cdots,K+M\}$ with $\mathcal D \setminus \mathcal S = \emptyset$, where $\mathcal{R}_{(\mathcal{S},\mbox{\scriptsize \boldmath $y$})}$ is defined as in (\ref{CompoundCriterionAppendix}).

Compared with the proof of Theorem \ref{Theorem3}, the key difference is that, in this proof, we need to consider user subsets $\mathcal{S}$ with $\mathcal D \setminus \mathcal S = \emptyset$. In the rest of the proof, we will skip the discussions involving user subsets $\mathcal{S}$ with $\mathcal D \setminus \mathcal S \ne \emptyset$ since they are exactly the same as those in Appendix \ref{ProofTheorem3}.

Given a user subset $\mathcal{S}$ with $\mathcal D \setminus \mathcal S = \emptyset$, we define the following probability terms.

First, assume that $\mbox{\boldmath$w$}_{\mathcal D}$ is the transmitted message vector of users in $\mathcal D$, and $\mbox{\boldmath$g$}$ is the actual code index vector with $\mbox{\boldmath$g$} \in \mathcal{R}_{\mathcal D}$. Let $P_{t[\mbox{\scriptsize\boldmath$g$},\mathcal{S}]}$ be the probability that the weighted likelihood of the transmitted codeword vector is no larger than the corresponding typicality threshold, as defined in (\ref{ProofPtCC}).

Second, assume that $\tilde{\mbox{\boldmath$w$}}_{\mathcal D}$ is the transmitted message vector of users in $\mathcal D$, and $\tilde{\mbox{\boldmath$g$}}$ is the actual code index vector, with $\tilde{\mbox{\boldmath$g$}} \notin \mathcal{R}_{\mathcal D}\cup \widehat{\mathcal{R}}_{\mathcal D}$. Let $P_{i[\tilde{\mbox{\scriptsize\boldmath$g$}},\mbox{\scriptsize\boldmath$g$},\mathcal{S}]}$ be the probability that the decoder finds a codeword $(\mbox{\boldmath$w$}_{\mathcal D},\mbox{\boldmath$g$})$ with $(\mbox{\boldmath$w$}_{\mathcal D},\mbox{\boldmath$g$}) \stackrel{\mathcal S}{=} (\tilde{\mbox{\boldmath$w$}}_{\mathcal D},\tilde{\mbox{\boldmath$g$}})$ and $\mbox{\boldmath$g$} \in \mathcal{R}_{\mathcal D}$, such that its weighted likelihood is larger than the corresponding typicality threshold, as defined in (\ref{ProofPiCC}).

With the probability definitions, by applying the union bound over all $\mathcal{S}$, we upper-bound $\mbox{GEP}_{\mathcal D}(\alpha)$ by
\begin{eqnarray}
\label{ProofPes2}
&& \mbox{GEP}_{\mathcal D}(\alpha) \le \frac{1}{\sum_{\mbox{\scriptsize\boldmath$g$}}\exp\left(-N\alpha({\mbox{\boldmath$g$}})\right)}\Biggl\{   \nonumber \\
&& \sum_{\mbox{\scriptsize\boldmath$g$}\in \mathcal{R}_{\mathcal D}} \Biggl[  \sum_{\tiny \mathcal{S}\subset \{1,\cdots,K+M\}} P_{t[\mbox{\scriptsize\boldmath$g$},\mathcal{S}]}e^{-N\alpha(\mbox{\scriptsize\boldmath$g$})}   \nonumber \\
&& + \sum_{\tiny \begin{array}{c}\mathcal{S}\subset \{1,\cdots,K+M\} \\ \mathcal{D}\setminus\mathcal{S}\ne \emptyset \end{array}} \sum_{\tilde{\mbox{\scriptsize\boldmath$g$}} \in \mathcal{R}_{\mathcal D},\tilde{\mbox{\scriptsize\boldmath$g$}}_{\mathcal{S}} = \mbox{\scriptsize\boldmath$g$}_{\mathcal{S}}} P_{m[\mbox{\scriptsize\boldmath$g$},\tilde{\mbox{\scriptsize\boldmath$g$}},\mathcal{S}]}e^{-N\alpha(\mbox{\scriptsize\boldmath$g$})}\Biggl] \nonumber \\
&& \left. +\sum_{\tilde{\mbox{\scriptsize\boldmath$g$}} \in \widehat{\mathcal{R}}_{\mathcal D}} \sum_{\tiny \begin{array}{c}\mathcal{S}\subset \{1,\cdots,K+M\} \\ \mathcal{D}\setminus\mathcal{S}\ne \emptyset \end{array}}
\sum_{\mbox{\scriptsize\boldmath$g$} \in \mathcal{R}_{\mathcal D},\mbox{\scriptsize\boldmath$g$}_{\mathcal{S}} = \tilde{\mbox{\scriptsize\boldmath$g$}}_{\mathcal{S}}} P_{i[\tilde{\mbox{\scriptsize\boldmath$g$}},\mbox{\scriptsize\boldmath$g$},\mathcal{S}]}e^{-N\alpha(\tilde{\mbox{\scriptsize\boldmath$g$}})} \right.    \nonumber \\
&&  +\sum_{\tilde{\mbox{\scriptsize\boldmath$g$}} \not\in \mathcal{R}_{\mathcal D}\cup \widehat{\mathcal{R}}_{\mathcal D}}  \sum_{\tiny \mathcal{S}\subset \{1,\cdots,K+M\}} \nonumber \\
&& \left. \sum_{\mbox{\scriptsize\boldmath$g$} \in \mathcal{R}_{\mathcal D},\mbox{\scriptsize\boldmath$g$}_{\mathcal{S}} = \tilde{\mbox{\scriptsize\boldmath$g$}}_{\mathcal{S}}} P_{i[\tilde{\mbox{\scriptsize\boldmath$g$}},\mbox{\scriptsize\boldmath$g$},\mathcal{S}]}e^{-N\alpha(\tilde{\mbox{\scriptsize\boldmath$g$}})} \right\}.
\end{eqnarray}
Next, we will derive individual upper-bounds for each of the terms on the right hand side of (\ref{ProofPes2}), under the assumption that $\mathcal{S}$ satisfies $\mathcal D \setminus \mathcal S = \emptyset$.

Note that $P_{m[\mbox{\scriptsize\boldmath$g$},\tilde{\mbox{\scriptsize\boldmath$g$}},\mathcal{S}]}e^{-N\alpha(\mbox{\scriptsize\boldmath$g$})}$ is not involved in terms with $\mathcal{S}$ and $\mathcal D \setminus \mathcal S = \emptyset$.

As shown in Step II of Appendix \ref{ProofTheorem3}, given that $\mbox{\boldmath$g$}\in \mathcal{R}_{\mathcal D}$, for any $s_1>0$, we can upper-bound $P_{t[\mbox{\scriptsize\boldmath$g$},\mathcal{S}]}e^{-N\alpha({\mbox{\boldmath$g$}})} $ by (\ref{MCBound4P_t}).

Following a similar derivation in Step III of Appendix \ref{ProofTheorem3}, given $\tilde{\mbox{\boldmath$g$}} \notin \mathcal{R}_{\mathcal D}\cup \widehat{\mathcal{R}}_{\mathcal D}$ and $\mbox{\boldmath$g$} \in \mathcal{R}_{\mathcal D}$, for any $s_2 >0$ and $\tilde{\rho} > 0$, we can upper-bound $P_{i[\tilde{\mbox{\scriptsize\boldmath$g$}},\mbox{\scriptsize\boldmath$g$},\mathcal{S}]}e^{-N\alpha(\tilde{\mbox{\scriptsize\boldmath$g$}})}$ by
\begin{eqnarray}
\label{MCBound4P_i2}
&& P_{i[\tilde{\mbox{\scriptsize\boldmath$g$}},\mbox{\scriptsize\boldmath$g$},\mathcal{S}]}e^{-N\alpha(\tilde{\mbox{\scriptsize\boldmath$g$}})} \le \max_{\mbox{\scriptsize\boldmath$g$}' \notin \mathcal{R}_{\mathcal D}\cup \widehat{\mathcal{R}}_{\mathcal D},\mbox{\scriptsize\boldmath$g$}'_{\mathcal{S}} = \mbox{\scriptsize\boldmath$g$}_{\mathcal{S}}}   e^{-N\alpha(\mbox{\scriptsize\boldmath$g$}')}     \nonumber \\
&& \times \sum_{\mbox{\scriptsize\boldmath$y$}}   E_{\mbox{\scriptsize\boldmath$\theta$}_{\mathcal{S}\cap \mathcal D}} \Bigl[  E_{\mbox{\scriptsize\boldmath$\theta$}_{\mathcal D \setminus \mathcal{S}}} \left[ P(\mbox{\boldmath$y$}|\mbox{\boldmath$x$}_{(\mbox{\scriptsize\boldmath$w$}'_{\mathcal D},\mbox{\scriptsize\boldmath$g$}'_{\mathcal D})},\mbox{\boldmath$g$}'_{\bar{\mathcal D}}) \right]  \nonumber\\
&& \quad \times \left\{ E_{\mbox{\scriptsize\boldmath$\theta$}_{\mathcal D \setminus \mathcal{S}}} \left[ P(\mbox{\boldmath$y$}|\mbox{\boldmath$x$}_{(\mbox{\scriptsize\boldmath$w$}_{\mathcal D},\mbox{\scriptsize\boldmath$g$}_{\mathcal D})},\mbox{\boldmath$g$}_{\bar{\mathcal D}})^{\frac{s_2}{\tilde{\rho}}} \right] \right\}^{\tilde{\rho}}      \nonumber \\
&& \quad \times e^{Ns_2 (\tau_{(\mbox{\tiny\boldmath$g$},\mathcal{S})}(\mbox{\scriptsize\boldmath$x$}_{{\mathcal S} \cap {\mathcal D} }, \mbox{\scriptsize\boldmath$y$})-\alpha(\mbox{\scriptsize\boldmath$g$})) }   \left. e^{N\tilde{\rho}\sum_{k\in\mathcal D\setminus\mathcal{S}}r_k} \right].
\end{eqnarray}

Let $\tilde{\mbox{\boldmath$g$}}^*\not\in\mathcal R_{\mathcal D}\cup \widehat{\mathcal{R}}_{\mathcal D}$ be the code index vector that maximizes the right hand side of (\ref{MCBound4P_i2}). By following the same derivation from (\ref{TauEquality}) to (\ref{MCBound4P_iB}), we get
\begin{eqnarray}
\label{BoundPiPt2}
&& P_{t[\mbox{\scriptsize\boldmath$g$},\mathcal{S}]}e^{-N\alpha(\mbox{\scriptsize\boldmath$g$})} \nonumber \\
&& \quad \le \max_{\tiny \begin{array}{c} \mbox{\scriptsize\boldmath$g$}' \notin \mathcal{R}_{\mathcal D}\cup \widehat{\mathcal{R}}_{\mathcal D} \\ \mbox{\scriptsize\boldmath$g$}'_{\mathcal{S}} = \mbox{\scriptsize\boldmath$g$}_{\mathcal{S}}\end{array}} \exp \left\{ -NE_{i \mathcal{D}}(\mathcal{S}, \mbox{\boldmath$g$},\mbox{\boldmath $g$}') \right\}, \nonumber \\
&& P_{i[\tilde{\mbox{\scriptsize\boldmath$g$}},\mbox{\scriptsize\boldmath$g$},\mathcal{S}]}e^{-N\alpha(\tilde{\mbox{\scriptsize\boldmath$g$}})} \nonumber \\
&& \quad \le \max_{\tiny \begin{array}{c} \mbox{\scriptsize\boldmath$g$}' \notin \mathcal{R}_{\mathcal D}\cup \widehat{\mathcal{R}}_{\mathcal D} \\ \mbox{\scriptsize\boldmath$g$}'_{\mathcal{S}} = \mbox{\scriptsize\boldmath$g$}_{\mathcal{S}}\end{array}} \exp \left\{ -NE_{i \mathcal{D}}(\mathcal{S}, \mbox{\boldmath$g$},\mbox{\boldmath $g$}') \right\},
\end{eqnarray}
where $E_{i \mathcal{D}}(\mathcal{S}, \mbox{\boldmath$g$},\mbox{\boldmath $g$}') $ is given in (\ref{EmEiDDcoder}).

Combining the results involving all user subset $\mathcal{S}$, we obtain (\ref{SRSBound2}).

\end{proof}

\subsection{Proof of Theorem \ref{Theorem9}}
\label{ProofTheorem9}
\begin{proof}
Given $\mbox{\boldmath $g$}\in C$ being the actual code index vector and $\hat{\mbox{\boldmath $g$}}$ being its estimate at the receiver, we have
\begin{equation}
Pr\{\hat{\mbox{\boldmath $g$}}\not\in C\}\le \sum_{\tilde{\mbox{\scriptsize \boldmath $g$}}\not\in C}E\left[\sum_{\mbox{\scriptsize \boldmath $y$}}P(\mbox{\boldmath $y$}|\mbox{\boldmath $g$})\phi_{[\mbox{\scriptsize \boldmath $g$}, \tilde{\mbox{\scriptsize \boldmath $g$}}]}(\mbox{\boldmath $y$})\right],
\end{equation}
where $\phi_{[\mbox{\scriptsize \boldmath $g$}, \tilde{\mbox{\scriptsize \boldmath $g$}}]}(\mbox{\boldmath $y$})$ is an indicator function with $\phi_{[\mbox{\scriptsize \boldmath $g$}, \tilde{\mbox{\scriptsize \boldmath $g$}}]}(\mbox{\boldmath $y$})=1$ if $P(\mbox{\boldmath $y$}|\tilde{\mbox{\boldmath $g$}})e^{-N\alpha(\tilde{\mbox{\scriptsize\boldmath$g$}})}> P(\mbox{\boldmath $y$}|\mbox{\boldmath $g$})e^{-N\alpha(\mbox{\scriptsize\boldmath$g$})}$ and $\phi_{[\mbox{\scriptsize \boldmath $g$}, \tilde{\mbox{\scriptsize \boldmath $g$}}]}(\mbox{\boldmath $y$})=0$ otherwise.

Following a similar bounding approach presented in Appendix \ref{ProofTheorem3}, with $0<s\le 1$, we can upper bound $E\left[\sum_{\mbox{\scriptsize \boldmath $y$}}P(\mbox{\boldmath $y$}|\mbox{\boldmath $g$})\phi_{[\mbox{\scriptsize \boldmath $g$}, \tilde{\mbox{\scriptsize \boldmath $g$}}]}(\mbox{\boldmath $y$})\right]e^{-N\alpha(\mbox{\scriptsize\boldmath$g$})}$ by
\begin{eqnarray}
&& E\left[\sum_{\mbox{\scriptsize \boldmath $y$}}P(\mbox{\boldmath $y$}|\mbox{\boldmath $g$})\phi_{[\mbox{\scriptsize \boldmath $g$}, \tilde{\mbox{\scriptsize \boldmath $g$}}]}(\mbox{\boldmath $y$})\right]e^{-N\alpha(\mbox{\scriptsize\boldmath$g$})} \nonumber \\
&& \le  E\left[\sum_{\mbox{\scriptsize \boldmath $y$}}\left[P(\mbox{\boldmath $y$}|\mbox{\boldmath $g$})e^{-N\alpha(\mbox{\scriptsize\boldmath$g$})}\right]^{s}\left[P(\mbox{\boldmath $y$}|\tilde{\mbox{\boldmath $g$}})e^{-N\alpha(\tilde{\mbox{\scriptsize\boldmath$g$}})}\right]^{(1-s)} \right]   \nonumber\\
&&        =\exp\Biggl(-N\log\sum_Y \left[P(Y|\mbox{\boldmath $g$})e^{-N\alpha(\mbox{\scriptsize\boldmath$g$})}\right]^{s} \nonumber \\
&& \quad \times\left[P(Y|\tilde{\mbox{\boldmath $g$}})e^{-N\alpha(\tilde{\mbox{\scriptsize\boldmath$g$}})}\right]^{(1-s)}\Biggr).
\end{eqnarray}
The bound given in (\ref{CommParameterDetectionBound}) then follows.
\end{proof}



\end{document}